\documentclass[prl,twocolumn,preprintnumbers,amsmath,amssymb,superscriptaddress,nolongbibliography]{revtex4-1}
\usepackage{graphicx}
\usepackage{dcolumn}
\usepackage{bm}

\usepackage{color}
\usepackage[normalem]{ulem}
\usepackage{hyperref}

\definecolor{strawberry}{rgb}{1.0,0.0,0.5}

\begin{document}

\title{Structural signatures of ultrastability in a deposited glassformer}

\author{Fabio Leoni}
\affiliation{Dipartimento di Fisica, Universit\`a degli Studi di Roma La Sapienza, Piazzale Aldo Moro 5, Rome, 00185, Italy}
\author{Fausto Martelli}
\affiliation{IBM Research Europe, Hartree Centre, Daresbury WA4 4AD, UK}
\author{C. Patrick Royall}
\affiliation{Gulliver UMR CNRS 7083, ESPCI Paris, Universit\'e PSL, 75005 Paris, France.}
\affiliation{School of Chemistry, Cantock’s Close, University of Bristol, BS8 1TS, UK}
\affiliation{Centre for Nanoscience and Quantum Information, Tyndall Avenue, Bristol BS8 1FD,
UK}
\affiliation{H.H. Wills Physics Laboratory, Tyndall Ave., Bristol, BS8 1TL, UK}
\author{John Russo}
\affiliation{Dipartimento di Fisica, Universit\`a degli Studi di Roma La Sapienza, Piazzale Aldo Moro 5, Rome, 00185, Italy}

\begin{abstract}
Glasses obtained from vapor deposition on a cold substrate
have superior thermodynamic and kinetic stability with respect to
ordinary glasses.
Here we perform molecular dynamics simulations of vapor deposition of a model glass-former and investigate the origin of its high stability compared to that of ordinary glasses. We find that the vapor deposited glass is characterized by locally favoured structures (LFS) whose occurrence correlates with its stability, reaching a maximum at the optimal deposition temperature. The formation of LFS is enhanced near the free surface, hence supporting the idea that the stability of vapor deposited glasses is connected to the relaxation dynamics at the surface.
\end{abstract}


\maketitle

The formation of glass films obtained from vapor deposition on a substrate is attracting ever increasing attention in the scientific community due to its technological and theoretical implications \cite{Ediger2017,Rodriguez2022,rossnagel2003thin,rossnagel1999sputter}.
Vapor deposited glasses (DG) obtained at specific conditions can show remarkable properties \cite{Swallen2007}, which for a conventional glass (e.g., obtained by quenching of a liquid melt) would require  thousands of years of aging \cite{Lyubimov2013}.
These glasess are called ultra-stable glasses (USG) and are obtained by careful consideration of the substrate temperature ($T_{sub}$) kept below the glass transition temperature ($T_g$), and of the deposition rate $\gamma_{DG}$ \cite{Swallen2007,Ediger2017,Ramos2011,Kearns2007,Leon2008,Berthier2017,Luo2018}.
While conventional glasses show low mechanical stability which results in spontaneous aging and devitrification over time \cite{Angell2000} leading to structural changes during its lifetime in many applications, USG exhibit enhanced kinetic and thermodynamic stability with respect to them.
Understanding these properties has the potential to bring fundamental insights into the nature of deeply supercooled liquids and structural properties of amorphous solids from one side, and to provide new routes for engineering materials with specific properties from the other side, for example in the realization of OLED displays \cite{Ediger2017}, in the stabilization of amorphous pharmaceuticals \cite{Ediger2017}, in the miniaturization of next-generation computing components and interconnect technologies~\cite{simon2020role,sil2021impact,jamison2015sio,rosenberg2000copper,nogami2022advanced}, or in improving coatings and design of composite materials using metallic glasses \cite{Luo2018,Ashby2006,wang2009}.

The enhanced properties of USG with respect to conventional glasses originate from structural configurations corresponding to deeper minima in the potential energy landscape (PEL) \cite{GUPTA2019}.
The PEL is strictly related to dynamical, thermodynamic, structural, and chemical properties of amorphous materials \cite{Ediger2017,Qiu2016,Dawson2009}.

Experiments \cite{Swallen2007,Bell2003,Ellison2003}, simulations \cite{Lyubimov2013,reid2016age,Berthier2017} and theory \cite{stevenson2008} suggest that the key to understand the formation of USG by vapor deposition is the enhanced mobility at the surface of the deposited layer, which correlates with the stability of the glass. Indeed, it is estimated \cite{Swallen2007} that, up to temperatures far below $T_g$, molecules at the surface remain mobile for several bulk relaxation times during which, for suitable control parameters, they find near-equilibrium configurations before being buried by other molecules undergoing deposition.
However, it remains unclear how exactly the enhanced mobility at the surface plays its role in the formation of ultra-stable deposited glasses \cite{Samanta2019}.

\begin{figure}[!t] 
\begin{center}
\includegraphics[width=8.5cm]{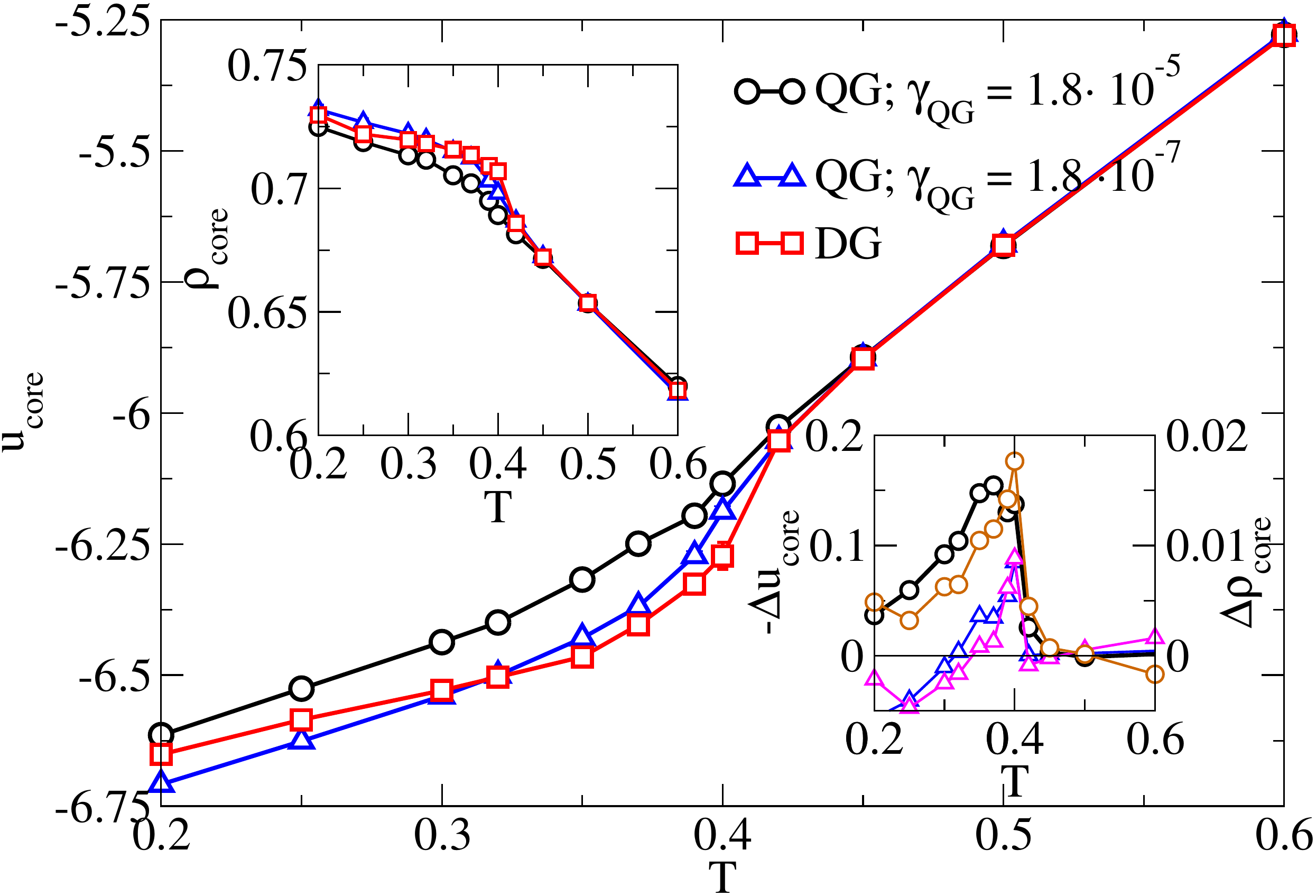}
\caption{\label{fig:u_vs_T} Potential energy per particle $u_{core}$ versus temperature $T$ for the DG and QG at annealing rates $\gamma_{QG}=1.8\cdot 10^{-5}, 1.8\cdot 10^{-7}$). Left-top inset: density versus temperature associated with the core of the DG and QG. Right-bottom inset: $u_{core}$ difference between QG and DG, and $\rho_{core}$ difference between DG and QG for $\gamma_{QG}=1.8\cdot 10^{-5}$.}
\end{center}
\end{figure}

\begin{figure}[!t] 
\begin{center}
\includegraphics[width=8.5cm]{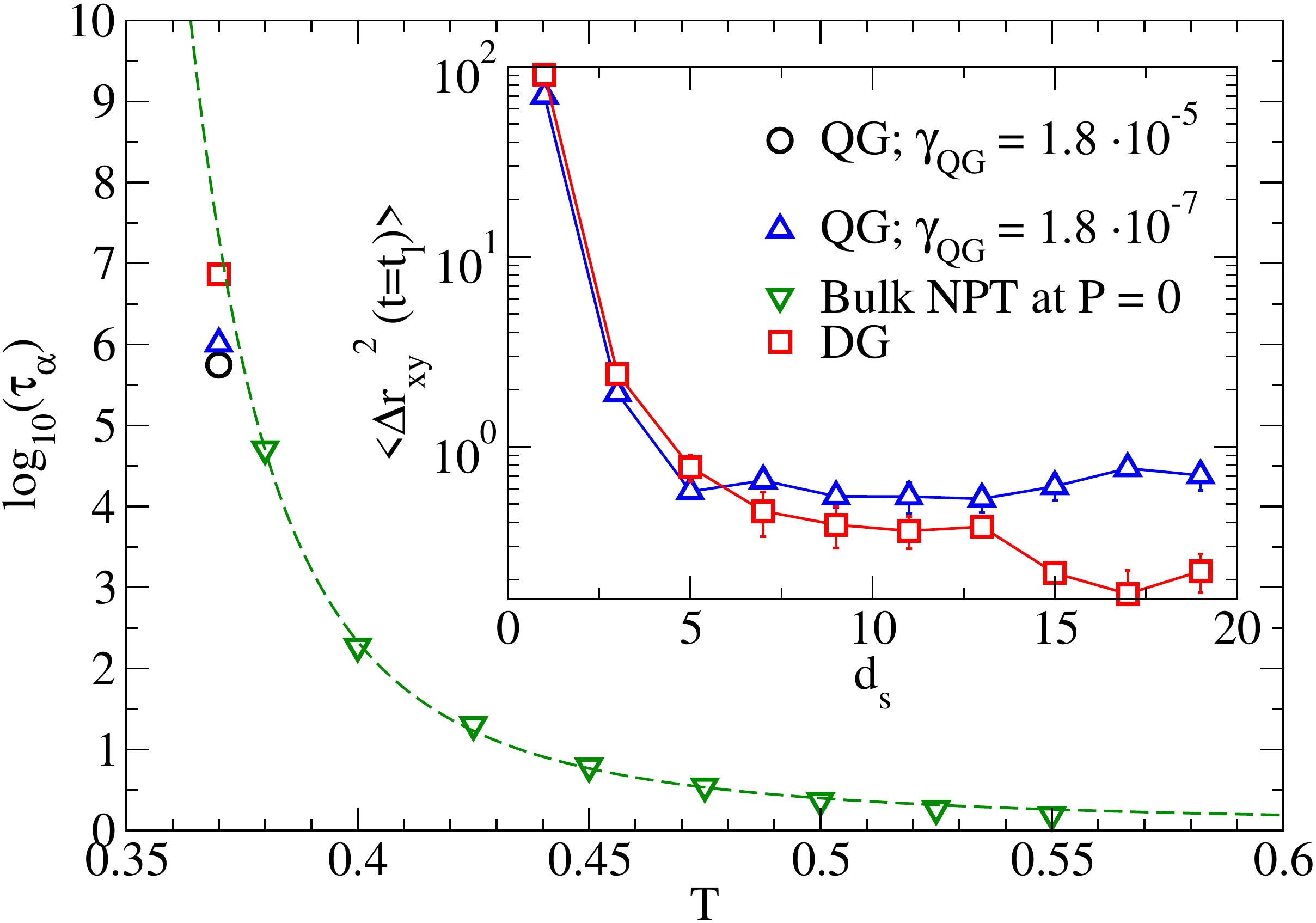}
\caption{\label{fig:alpha} Structural relaxation time versus temperature for the bulk at $P=0$ (green triangles) fitted with the VFT law (dashed line, see Supplementary).
The DG (red square) and QG at $\gamma_{QG}=1.8\cdot 10^{-5}$ (black circle) and $\gamma_{QG}=1.8\cdot 10^{-7}$ (blue triangle) are shown for $T=0.37$. Inset: MSD in the $xy$ plane at the lag time $t_l=2\cdot 10^3$ as a function of the distance from the free surface $d_s$ at $T=0.37$.}
\end{center}
\end{figure}

A relevant question is whether the glasses produced by vapor deposition share the same structural properties of glasses obtained by cooling~\cite{reid2016age}.
While long range properties can be inferred from two-point correlation functions~\cite{torquato2003local,xie2013hyperuniformity,martelli2017large}, investigation of short range structures is very limited~\cite{berthier2010critical,leocmach2013importance,royall2018race}. Instead, different higher 
order methods have been developed for this purpose~\cite{tanaka2019revealing}. Recently, bond-orientational order parameters have uncovered a link between liquid and glassy water \cite{Martelli2020} and, together with Voronoi polyhedra, have been correlated to the stability of deposited glasses\cite{singh2013ultrastable}, while non-trivial symmetries have been found in glassy water adopting a generic approach~\cite{martelli2018local,martelli2018searching}. In this work, we follow Ref.~\cite{Jenkinson2017} and refer to icosahedral motifs as locally favoured structures (LFS).
A strong correlation between the number of LFS and temperature or packing fraction has been found in systems as different as colloids \cite{Royall2008b,Leocmach2012} and metallic glasses \cite{Royall2015,singh2013ultrastable}.
In particular, previous studies on deposited glasses have found correlations between global (averaged over space) descriptors associated to the full sample and its stability \cite{Dalal2012,singh2013ultrastable,reid2016age}, whereas here, in addition to the global structural characterization through the system density (see Fig.~\ref{fig:u_vs_T}), we perform a local characterization through LFS which can be correlated to spatially-dependent properties of the sample such as its mobility profile (see Figs.~\ref{fig:alpha}, \ref{fig:msd}, \ref{fig:msd2}).

To understand the connection between the stability of deposited glasses and their structural properties, here we numerically investigate the process of formation of a model metallic glass former for which the LFS in equilibrium have been thoroughly investigated.
The study of metallic glass-formers is of great interest also from the experimental point of view for the unique mechanical and functional properties that these materials possess \cite{wang2009,Ashby2006}.

We adopt the Wahnstr\"om (WAHN) model~\cite{Wahnstrom1991}, which is composed of an equimolar additive bi-disperse Lennard-Jones (LJ) mixture (see Supplementary Material).
Several studies \cite{Jenkinson2017,Wahnstrom1991,Malins2013,Coslovich2007,Pedersen2010} 
have shown that for this system the LFS consist of particles arranged in the icosahedral geometry.
This model thus allows us to directly correlate the thermodynamic properties of the glass with the fraction of LFS, and investigate the connection between glasses prepared via different routes.

We study the system by performing molecular dynamics simulations (using LAMMPS \cite{LAMMPS}).
The units of energy, length and time are $\epsilon$, $\sigma$ and $t^*=(m_1\sigma^2/\epsilon)^{1/2}$, respectively. The integration time step is set to 0.005. Quantities are expressed in these reduced units.
We first estimate the glass transition temperature in the bulk. Glasses prepared in presence of a free surface equilibrate at ambient pressure and for which we obtain $T_g\simeq 0.36$ (See Supplementary Material). 

\begin{figure}[!t] 
\begin{center}
\includegraphics[width=2.8cm]{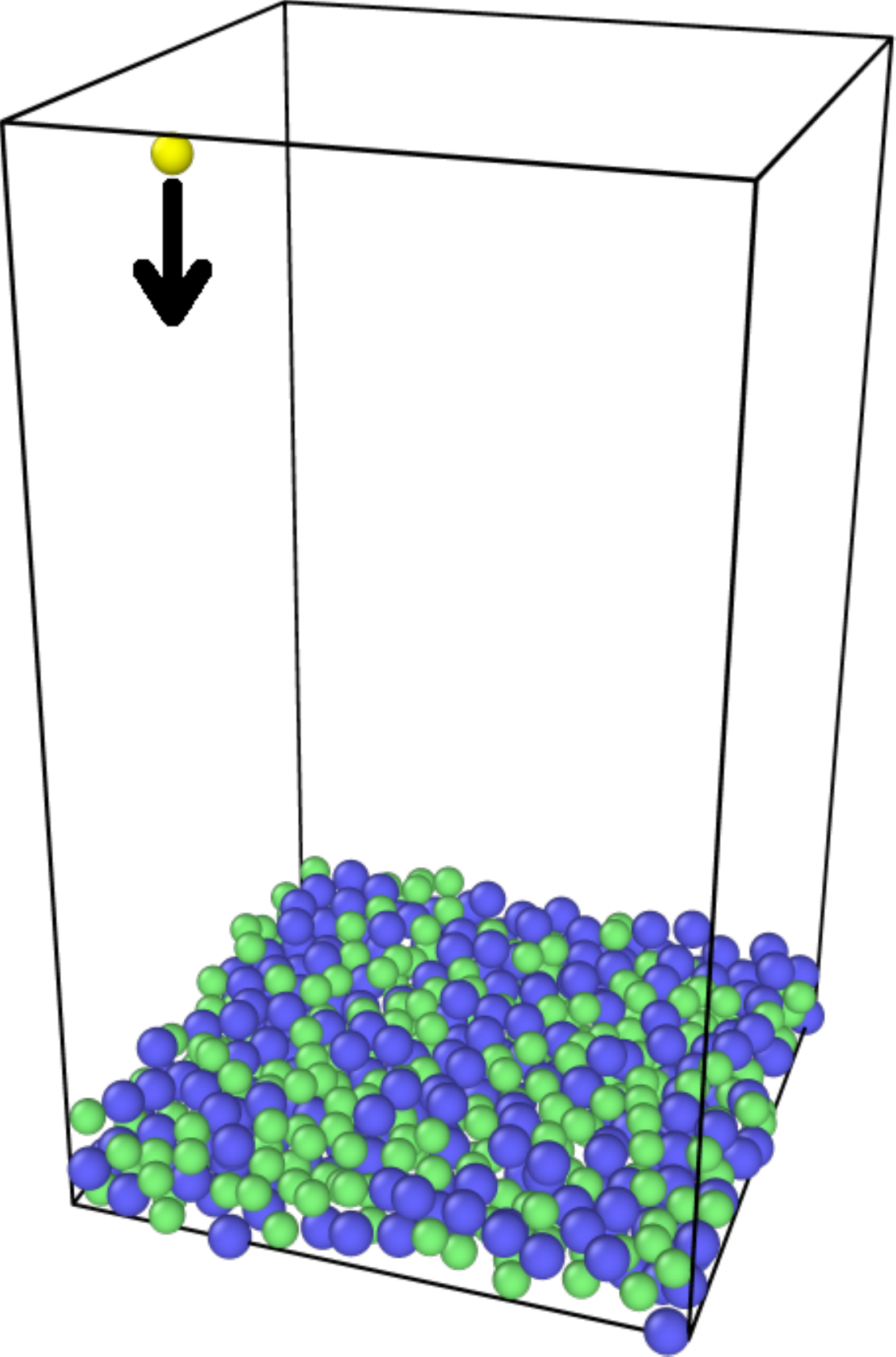}
\includegraphics[width=2.8cm]{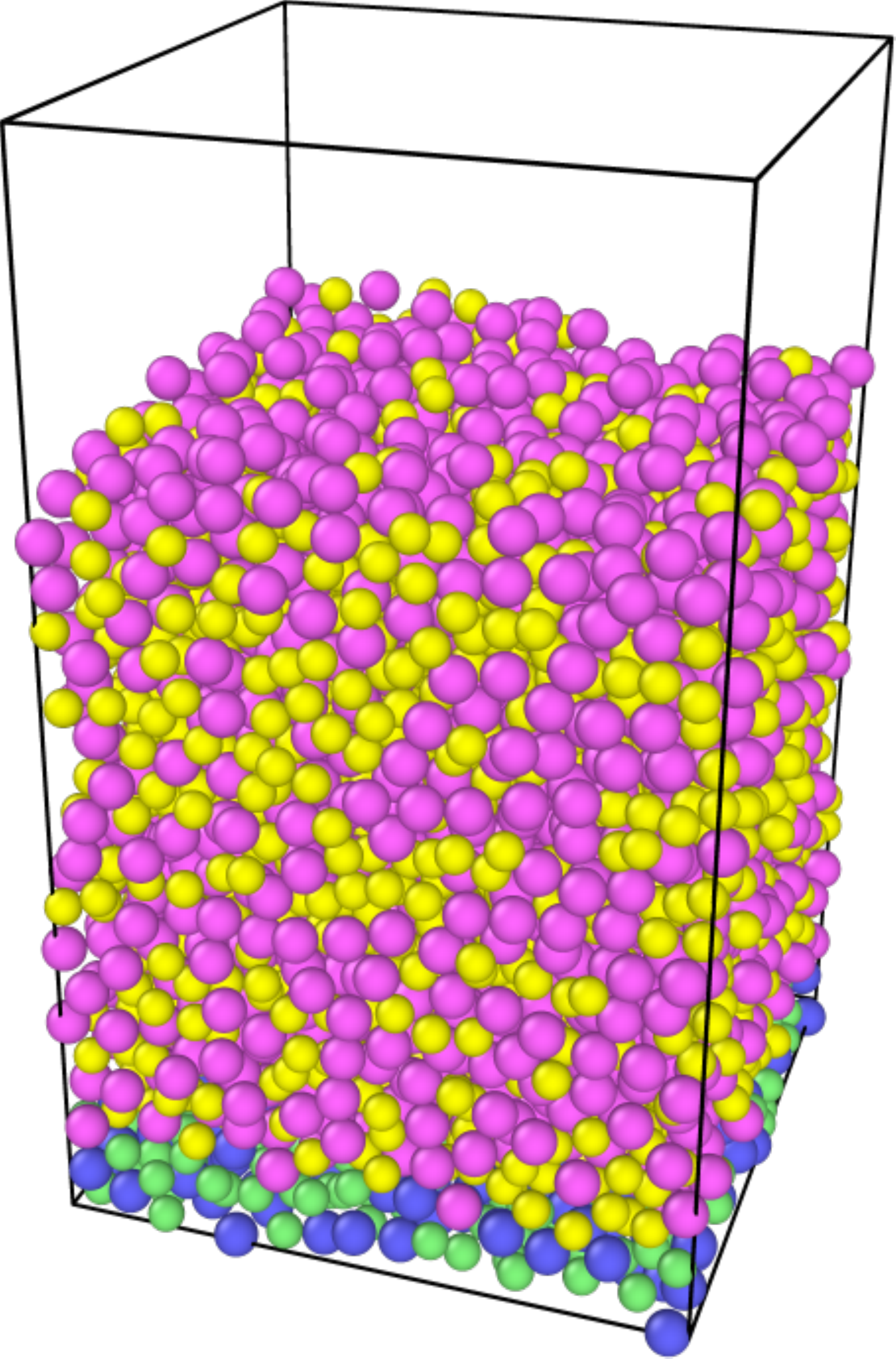}
\includegraphics[width=2.8cm]{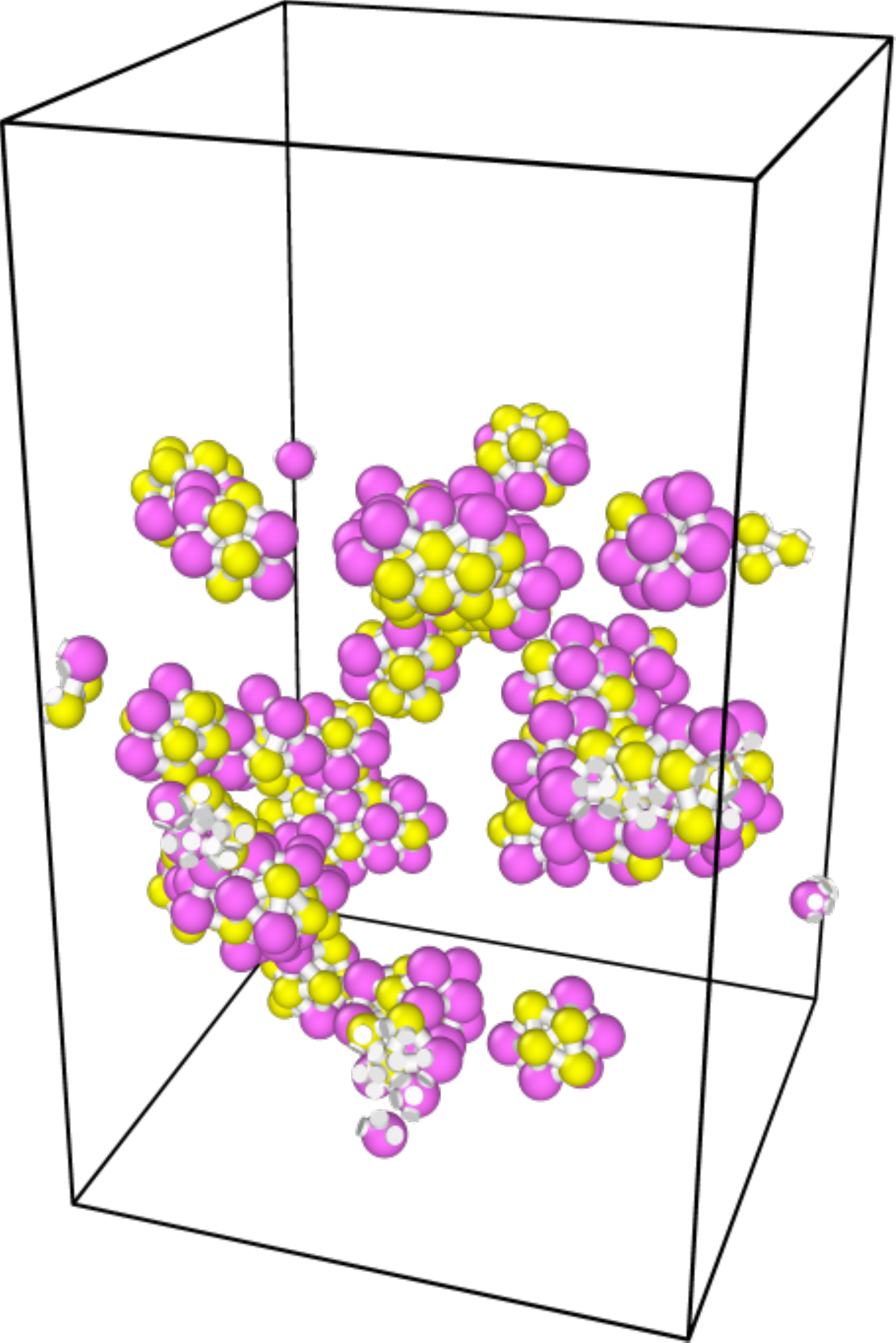}
\caption{\label{fig:snapshots} Snapshots showing for $T=0.42$ (left) the substrate formed by particles of type 1 (green) and 2 (blue) with the injection of a type 1 (yellow) particle, and (central) the deposited layer formed by particle of type 1 (yellow) and 2 (magenta). In the right panel particles belonging to icosahedra are shown. The snapshots are made with Ovito \cite{Stukowski2009}.}
\end{center}
\end{figure}

We compare both structural and dynamical properties of the DG with that of a conventional quenched glass (QG). 
The DG is obtained, similarly to Refs.~\cite{Lyubimov2013,Berthier2017},
by injecting a particle from a random position along x and y at the top of the simulation box with a frequency described by the deposition rate $\gamma_{DG}$ up to deposit a total of $N=4000$ particles, i.e. 2000 of type 1 and 2000 of type 2 (see Fig.~\ref{fig:snapshots}).
$\gamma_{DG}$ is given by the ratio between the thickness of the deposited layer and the elapsed time: $\gamma_{DG}=\Delta z/\Delta t$. We inject one particle every $5\cdot 10^4$ integration steps, so that we obtain $\gamma_{DG}\simeq 10^{-7}\sigma/dt=2\cdot 10^{-5}$.
The substrate (see Fig.~\ref{fig:snapshots}) placed at the bottom of the simulation box, periodic in $x$ and $y$, is composed of a disordered WAHN mixture with 250 particles of type 1 and 250 of type 2 at a number density in reduced units equals to $\rho^*=\rho(\sigma)^3=0.75$ forming a layer of thickness $\sim3\sigma$, and kept at temperature $T_{sub}$ (NVT ensemble, see Supplementary Material).

To prepare the QG with the same boundary conditions as the DG, we melt the DG by instantaneously heating it up to $T=1$, let the system equilibrate for a time of $5\cdot 10^3$, and then cool it down to $T=0.1$ at cooling rate $\gamma_{QG}$. In doing so we keep the substrate in contact with the deposited layer. The two cooling protocols we consider go from $T=1$ to $T=0.1$ in a time of $5\cdot 10^6$ and $5\cdot 10^4$, which correspond to $\gamma_{QG}=1.8\cdot 10^{-7}$ and $1.8\cdot 10^{-5}$, respectively.
To convert reduced units in real numbers we consider type 1 particles to be Argon atom with $\epsilon/k_B=120$~K, where $k_B$ is the Boltzmann constant, and $t^*=2.2$~ps \cite{pedersen2007,Pedersen2010}, such that $\gamma_{QG}\sim 0.01$~K/ns and $\gamma_{QG}\sim 1$~K/ns, respectively.
We choose a deposition rate such that the length of the simulations for DG (in a time of $10^6$) is intermediate between
the QG cooling times. In this way we can compare the structure of the as-deposited layer (immediately after deposition has finished) with that of quenched glasses that have had comparable relaxation times.
All the structural and dynamical properties that we consider are computed in the core of the deposited and quenched, unless otherwise specified. The core is defined as the part of the layer where particles have coordinates $z_{bot}+4\sigma<z<z_{top}-4\sigma$, with $z_{bot}$, $z_{top}$ the bottom and top edges of the layer, respectively (see Supplementary Material).
Among the different ways to quantify the stability of glasses \cite{Dalal2013,Fullerton2017,Rodriguez2022,Sepulveda2014,Dalal2012,Whitaker2015} we compute the core internal potential energy $u_{core}$ and density $\rho_{core}$.
Here we show (Fig.~\ref{fig:u_vs_T} and left-top inset therein) $u_{core}$ and $\rho_{core}$ versus $T$ for the deposited and quenched glasses at different cooling rates.
In the right-bottom inset of Fig.~\ref{fig:u_vs_T} we show both the opposite of the internal potential energy per particle difference $\Delta u_{core}$ (also shown in Fig.~\ref{fig:Delta_u_SM}) and density difference $\Delta\rho_{core}$ between DG and QG for both $\gamma_{QG}$. 
From it we obtain that the optimal temperature of the substrate to get a stable deposited glass falls in the range $T\in[0.32,0.42]$.

\begin{figure}[!t] 
\begin{center}
\includegraphics[width=8.5cm]{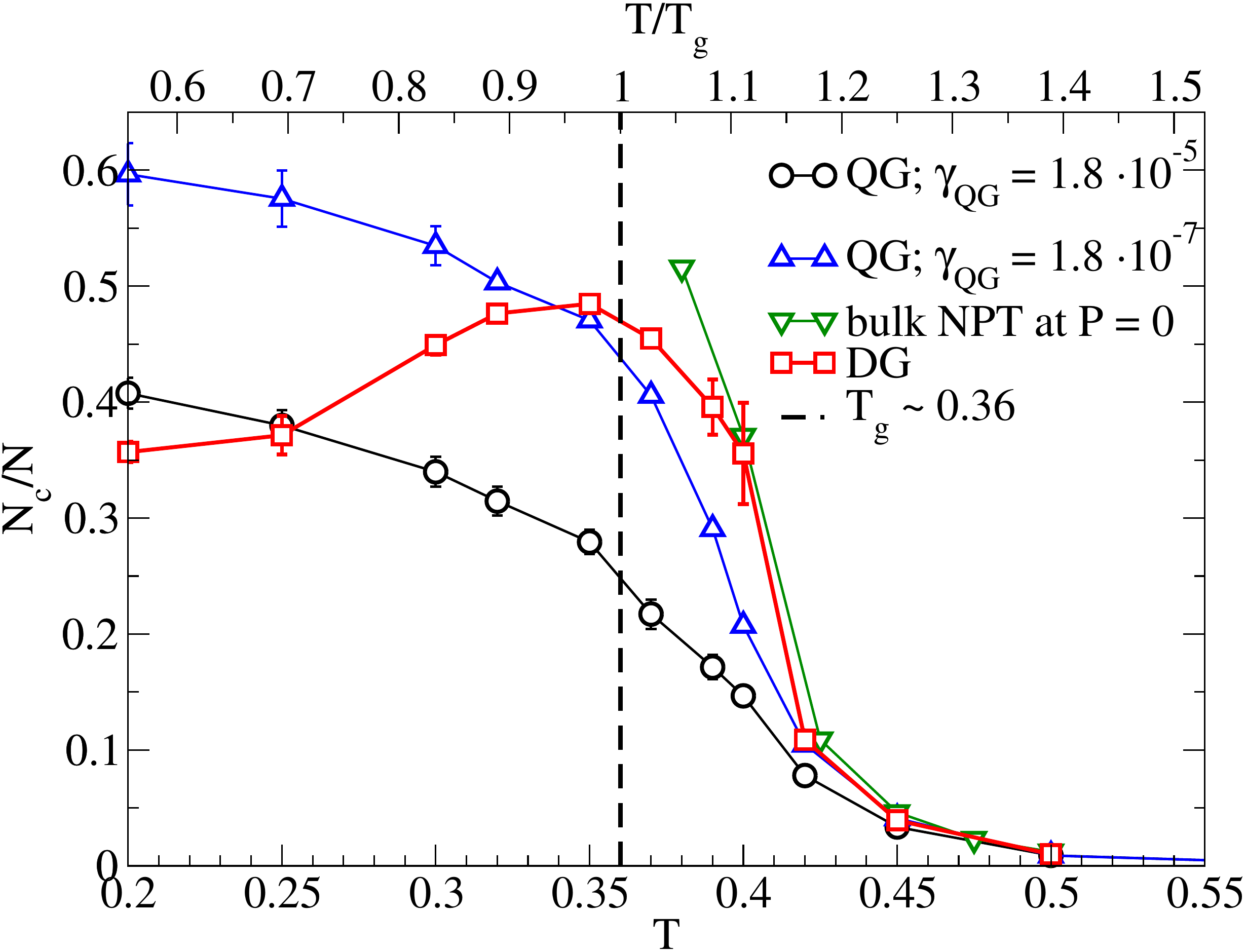}
\caption{\label{fig:TCC2} Fraction of particles detected within icosahedra with the TCC algorithm for DG (red squares), QG at two cooling rates (black circles and blue up-triangles), and equilibrated bulk at zero pressure (green down-triangles).}
\end{center}
\end{figure}

The LFS in Wahnstr\"om model is the icosahedron  \cite{Malins2013,Wahnstrom1991,Coslovich2007}.
To identify LFS we use the topological cluster classification (TCC) algorithm \cite{Malins2013}.
In Fig.~\ref{fig:TCC2} we show the average population of icosahedral clusters at different temperatures for DG (red squares) and QG at two cooling rates (black circles and blue up-triangles), and the equilibrated bulk at zero pressure (green down-triangles, down to the temperature where we can achieve equilibration). 
Interestingly, the deposited layer shows a peak in the distribution of icosahedra in the range of temperatures $0.32-0.36$ (corresponding to a fraction of $T_g$ in the range $0.85-1$), while annealed glasses show a continuous increase in the concentration of icosahedra for decreasing temperature for all cooling rates simulated. Despite the energy and density curves of the slow QG and DG glasses being very close in this region, the analysis of LFS shows a clear difference between the two glasses, and an increased stability of DG in a window of temperatures around $T_g$. Moreover, compared to QG with its monotonic temperature dependence, the DG has an excess of LFS only in this temperature range, showing that the effectiveness of vapor deposition depends on optimal glass forming conditions.
Comparing to the bulk $P=0$ simulations (green down-triangle points), we see that DG follows the equilibrium curve much closer than the QG samples. We thus find that the DG glass is characterized by the same LFS as the equilibrium case, being structurally equivalent to QG glasess before they fall out of equilibrium.
For larger values of $\gamma_{DG}$ we didn't observe any relevant shift in the temperature location of the peak of $\langle N_c/N\rangle$.
The main features of the glasses obtained from Fig.~\ref{fig:TCC2} do not change if, instead of the core of the layer, different regions are considered (see Supplementary Material).

We verified that also for another popular glassformer 
(the Kob-Andersen (KA) model 
\cite{Kob1994}) the deposited layer shows a peak in the distribution of the relevant LFS around an optimal temperature (see Supplementary Material).

%
%
While icosahedra capture short range order, Frank-Kasper (FK) polyhedra \cite{frank1958,Pedersen2010} allow to reveal the possible crystallization of the system into MgZn$_2$-like structures (given by a tetrahedral network of FK bonds). An FK analysis shows that these bonds are disordered and so no crystallization happens at the conditions considered in this work and the number of bonds follows a similar behavior to that of Fig.~\ref{fig:TCC2} (see Supplementary Material). 
We verified that, unlike molecular glasses made of anisotropic constituents \cite{Ediger2017}, the two-point correlation functions along different directions of DG and QG glasses, which are made of isotropic constituents, are isotropic (see Fig.\ref{fig:gr}), as observed also for the KA model in Ref.~\cite{singh2013ultrastable}.

\begin{figure}[!t] 
\begin{center}
\includegraphics[width=8.5cm]{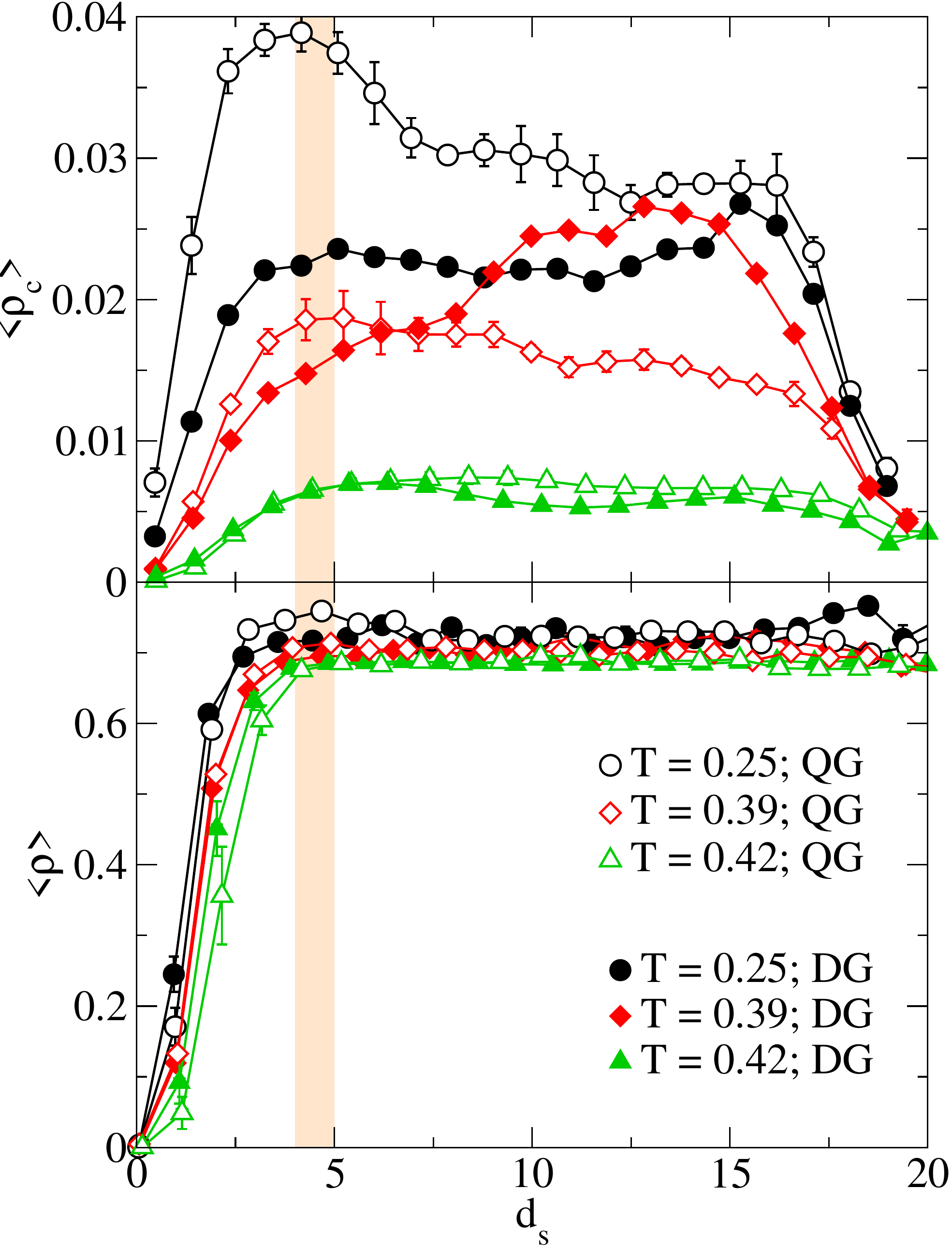}
\caption{\label{fig:n_c} Upper panel: average density profile $\langle\rho_c\rangle$ of the fraction of particles detected within icosahedra as a function of the distance from the free surface $d_s$. Lower panel: density profiles $\langle\rho\rangle$ versus $d_s$.}
\end{center}
\end{figure}

In the upper panel of Fig.~\ref{fig:n_c}, we show the average density profile, $\langle\rho_c\rangle$, of the fraction of particles detected within icosahedra as a function of the distance from the free surface, $d_s$, for the DG as-deposited, and the QG at $\gamma_{QG}=1.8\cdot 10^{-7}$, while in the lower panel we show the respective average number density profile, $\langle\rho\rangle$. At high temperature ($T=0.42$) the DG and QG show a similar, flat icosahedra profile, which mirrors the density profile, as expected at liquid-vapor interfaces \cite{godonoga2011}.
Decreasing temperature, the QG develops an increasingly pronounced peak in $\rho_c$ at a distance from the free surface $4\lesssim d_s\lesssim 5$ (range marked with a vertical bar in Fig.~\ref{fig:n_c}) and an overall increasing LFS density.
On the other hand, decreasing temperature the DG shows 
a peak at $13\lesssim d_s\lesssim 16$, while further decreasing temperature (below $T_g$) another peak develops at $4\lesssim d_s\lesssim 5$. The overall LFS density of the DG reaches a maximum around the temperature corresponding to the peak in the fraction of icosahedra (see Fig.~\ref{fig:TCC2}).
The behavior of the QG confirms the role of the free surface to enhance the formation of LFS (up to a distance $\sim 7\sigma$ from it) compared to the bulk. 
This same mechanism explains the behavior of the DG seen in Fig.~\ref{fig:n_c}: all particles of the as-deposited layer at some point during the deposition process were located at or near the free surface and consequently the concentration of LFS underneath the surface was enhanced. However, decreasing temperature the mobility near the free surface decreases (see Fig.~\ref{fig:msd}) so reducing the time available to particles near the free surface to find local stable configurations before getting buried by other particles undergoing deposition.

These results support the idea that the enhanced stability of vapor deposited glasses can be attributed to the faster relaxation of the surface~\cite{Moore2019}, which allows the formation of LFS for several layers below the surface. While the density profile is monotonic (inset of Fig.~\ref{fig:n_c}), the presence of peaks in the LFS profiles of Fig.~\ref{fig:n_c} shows that the surface enhances the formation of local structures compared to the bulk.

In Fig.~\ref{fig:alpha} we show the structural relaxation time $\tau_{\alpha}$ and the Vogel-Fulcher-Tammann (VFT) fit for the bulk simulations at $P=0$ (green down-triangles), together with the DG and the QG at both cooling rates for $T=0.37$, above $T_g$, but close to the maximum in the number of LFS (see Fig.~\ref{fig:TCC2}).
Relaxation times are obtained by fitting the self-intermediate scattering functions with the Kolrausch-Williams-Watts (KWW) law (see Supplementary Material) after the deposition or quenching process has ended. The figure shows that the DG (at the state point where LFS are close to the maximum) has a relaxation time closer to the bulk equilibrium extrapolated value (green dashed line) compared to the QG. The correspondence between the maximum in LFS (Fig.~\ref{fig:TCC2}) and the longest relaxation time (Fig.~\ref{fig:alpha}), and its closeness to the equilibrium extrapolated value, shows once more that vapor deposition produces  samples that are the closest to their bulk equilibrium counterparts and are considerably more aged than QG glasses.
In the inset of Fig.~\ref{fig:alpha} we compare the 2D mean square displacement (MSD) along the $xy$ plane at the lag time $t_l=2\cdot 10^3$ as a function of the distance from the free surface $d_s$ of the DG with the QG at $\gamma_{QG}=1.8\cdot 10^-7$.
Similarly to Ref.~\cite{zhang2022}, particles within the first $\sim 5\sigma$ layers underneath the free surface remain from 1 to almost 3 orders of magnitude faster than particles deep inside the layer, even at low $T$.
Furthermore, the MSD correlates with the stability of the glass (see Fig.~\ref{fig:msd}) and with the local structural classification (icosahedra are slower, see Fig.~\ref{fig:msd2}).

In conclusion, we have investigated the formation of a glass from vapor deposition for a well-known model glassformer 
(Wahnstr\"om mixture) for which the locally favoured structure is known and corresponds to icosahedral environments. This has allowed us to investigate the relation between glass stability and local structure formation, finding a close link between them. In particular, despite bulk quantities like energy and density being very similar between deposited and quenched glasses, the optimal deposition temperature occurs in correspondence to a peak in the fraction of LFS (Fig.~\ref{fig:TCC2}). Here, both structure and structural relaxation lie close to the extrapolated equilibrium value (obtained from bulk simulations at $P=0$), reinforcing the idea that the DG shares the same microscopic properties of well-aged glasses. We also investigated the role of the free surface in the formation of ultra-stable glasses, showing that at optimal deposition conditions the mobility profile along the deposition direction, enhanced at the surface and suppressed in the bulk, is correlated with the stability of the glass through the formation of an excess of LFS up to several layers below the surface.

\vspace{0.5cm}
\begin{acknowledgments}
FL and JR acknowledge support from the European Research Council Grant DLV-759187, and CINECA-ISCRA for HPC resources. CPR acknowledges funding through the ANR grant DiViNew.
\end{acknowledgments}


%

\clearpage
\newpage
\onecolumngrid

\begin{center}
{\bf\large{Supplementary Material for ``Structural and thermodynamic signature of ultrastability in a deposited glassformer''}}
\vspace{0.3cm}
\end{center}

\setcounter{equation}{0}
\setcounter{figure}{0}
\setcounter{table}{0}
\setcounter{section}{0}
\makeatletter
\renewcommand{\theequation}{S\arabic{equation}}
\renewcommand{\thefigure}{S\arabic{figure}}
\renewcommand{\thetable}{S\arabic{table}}
\renewcommand{\thesection}{S\arabic{section}}

\section{Wahnstr\"om mixture}

The Wahnstr\"om (WAHN) model~\cite{Wahnstrom1991} is composed of an equimolar additive bi-disperse Lennard-Jones (LJ) mixture with potential:
\begin{equation}
u_{LJ}(r)=4\epsilon_{\alpha\beta}\left[\left(\dfrac{\sigma_{\alpha\beta}}{r}\right)^{12}-\left(\dfrac{\sigma_{\alpha\beta}}{r}\right)^6\right],
\end{equation}
where $\alpha$ and $\beta$ can be particles of type 1 and 2. The energies are $\epsilon_{11}=\epsilon_{12}=\epsilon_{22}=\epsilon$, the masses $m_2/m_1=2$, and the diameters $\sigma_{22}/\sigma_{11}=1.2$ with the cross-interaction diameter given by the Lorentz-Berthelot mixing rule: $\sigma_{12}=(\sigma_{11}+\sigma_{22})/2$.
In the main text we set $\sigma_{11}=\sigma$.

\section{Deposition simulations details}

Particles of type 1 and 2 are introduced in the box alternately with a fixed vertical velocity $v_z=0.1$ and lateral components $v_{x,y}$ randomly extracted between -0.01 and 0.01. The top edge of the box elastically reflects particles that, eventually at high T, bounce or detach from the free surface.  
Injected particles are integrated in the NVE ensemble.
The substrate placed at the bottom of the simulation box (see Fig.~\ref{fig:snapshots}) is composed of a disordered WAHN mixture with 250 particles of type 1 and 250 of type 2 at a number density in reduced units equals to $\rho^*=\rho(\sigma)^3=0.75$, and kept at temperature $T_{sub}$ by using the Nos\'e-Hoover thermostat (NVT ensemble).
The configuration of the substrate is obtained by cutting a bulk system equilibrated at a temperature $T=0.4$ (in units of $\epsilon/k_B$, with $k_B$ the Boltzmann constant) and a number density in reduced units equal to $\rho^*=\rho(\sigma)^3=0.75$. 
We simulate three independently generated substrates to obtain averaged quantities.

The choice to take the substrate made of the same particles deposited on it avoids the possible formation of a gap between the deposited layer and the substrate \cite{Lyubimov2013}.
The lateral size of the box are $L_x=L_y=17.5\sigma$.
A spring force of stiffness $k=50$ is independently applied to each particle of the substrate to tether it to its initial position. This value of $k$ is large enough to keep particles fluctuating around their initial position and small enough to allow the equilibration of the deposited layer. As we verified by simulations, below $T_g$ the use of springs is not necessary to keep the particles of the substrate confined at the bottom of the box (in which case the bottom edge of the box is reflective as the top edge), while above $T_g$ they allow to avoid these particles to mix with deposited particles. 
The springs are always active, even during the heating and cooling processes to prepare the QG glass.

The value $T_g\simeq 0.36$ of the bulk system at $P=0$ is found from the intersection of the cooling curve in the glassy state and in the supercooled liquid state, which is remarkably in very good agreement with the $T_g$ corresponding to the value obtained from the empirical rule for the structural relaxation time $\tau_{\alpha}=100$~s in Argon units, as obtained from data in Fig.\ref{fig:alpha}, despite the separation of timescales between computer simulation and the timescale corresponding to $T_g$ in experimental systems.

\section{Structural relaxation time}

Short-time dynamics measurements employed to estimate long-time structural relaxation would be ideal when knowing a universal function connecting these two regimes \cite{Pogna2015}.
In the present case we study the kinetic stability of the glasses
computing the two-dimensional (2D, along x and y for DG and QG) and three-dimensional (3D, for the bulk) self-intermediate scattering function (ISF): $F(q,t)=\sum_{j=1}^{N}\exp\left[i\bm{q}\cdot(\bm{r}_j(t)-\bm{r}_j(0))\right]$,
where for the 2D case the sum runs over the $N=N_{core}$ particles that remain in the core (defined in the main text) up to times equal to or larger than $t$, and $\bm{r}_j(t)$ is the vector of coordinates $(x_j(t),y_j(t))$.
The wave vector is set to $|\bm{q}|=6$ for both cases, which corresponds to the first peak of the static structure factor. 

Estimating dynamic relaxation can be done by using different fitting curves of some correlation or response function \cite{Berthier2020}.  
Here we fit the ISF by using one of the common fitting functions \cite{Coslovich2007,Berthier2020}, i.e. the Kolrausch-Williams-Watts (KWW) law which gives the structural relaxation time $\tau_{\alpha}$: $F(q,t)=c\exp[-(t/\tau_{\alpha})^b]$, where $c$ is the parameter related to the height of the plateau before the alpha relaxation and $b$ describes the stretch of the exponential decay.
We fit the dependence on T of $\tau_{\alpha}$ with the Vogel-Fulcher-Tammann (VFT) law, suitable up to a characteristic temperature (see \cite{Coslovich2007}) which for the bulk at pressure $P=0$ is roughly $T=0.55$.   
Following VFT: $\tau_{\alpha}=\tau_{\infty}\exp[1/(K(T/T_0-1))]$. $K$ is related to the fragility of the glass, $T_0$ is the VFT divergence temperature.
The fitting parameters are: $T_0=0.258$, $K=0.4236$ and $\tau_{\infty}=0.0323$.

\section{Density profile}

\begin{figure*}[h] 
\begin{center}
\includegraphics[width=8cm]{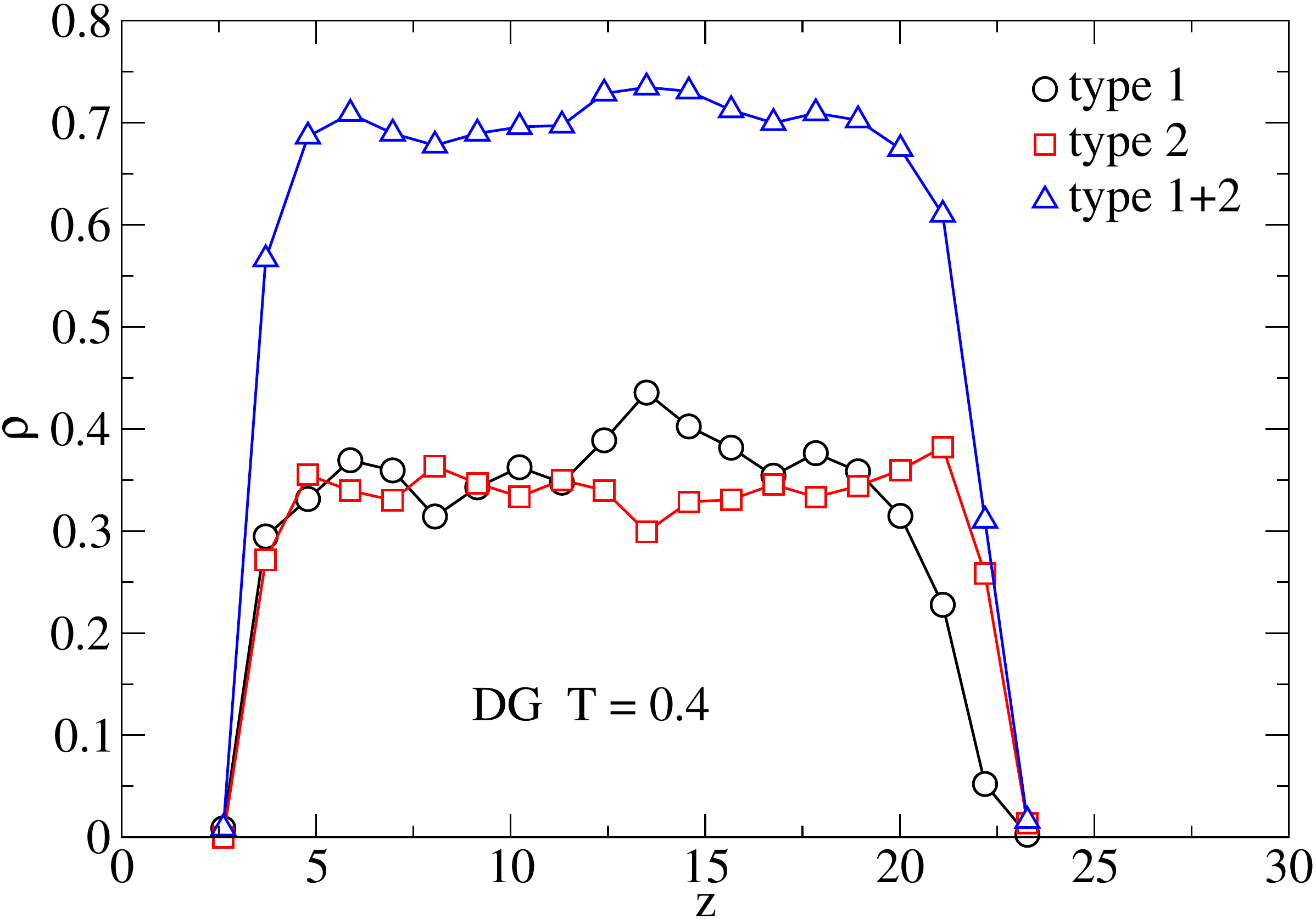}
\includegraphics[width=8cm]{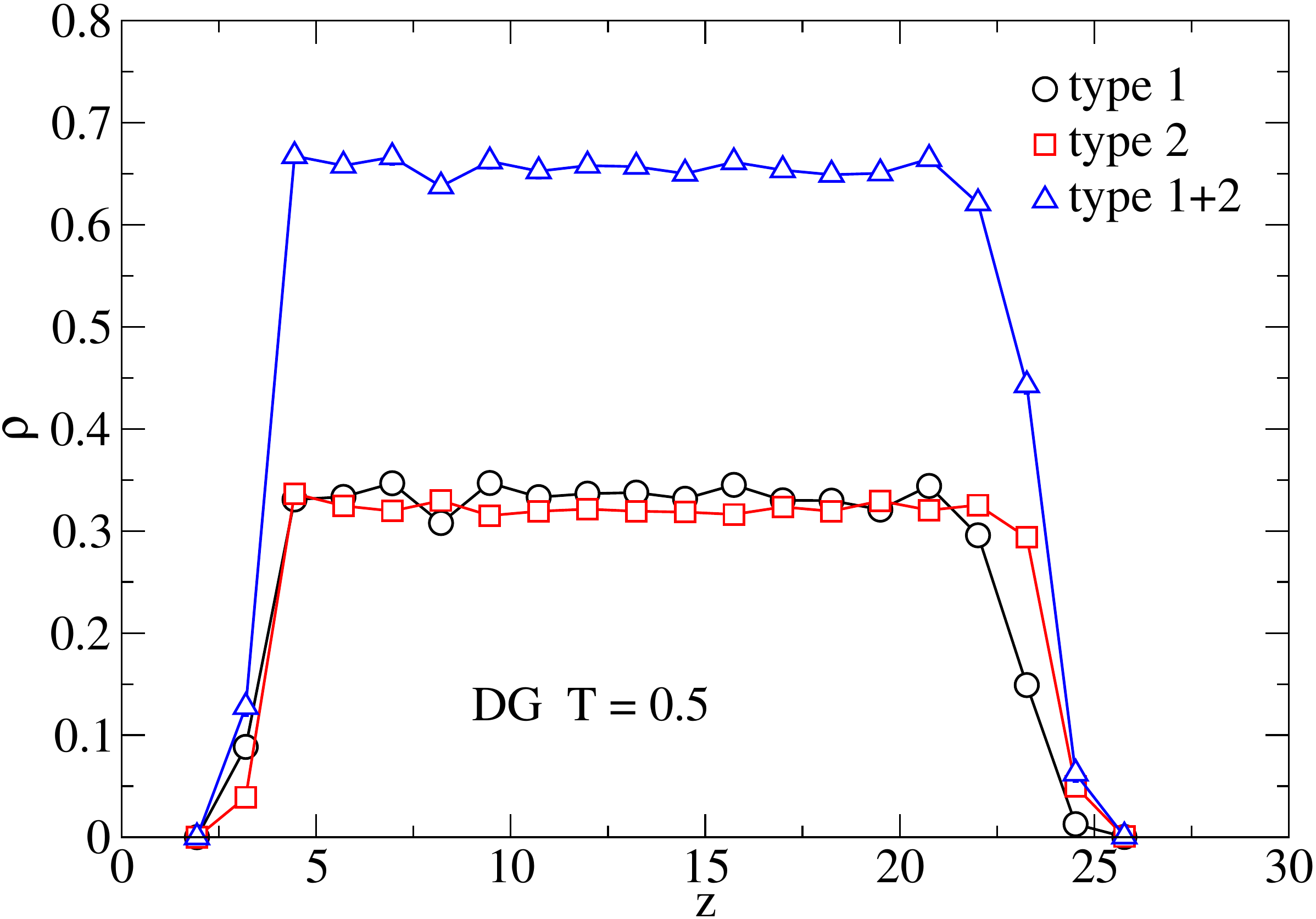}

\includegraphics[width=8cm]{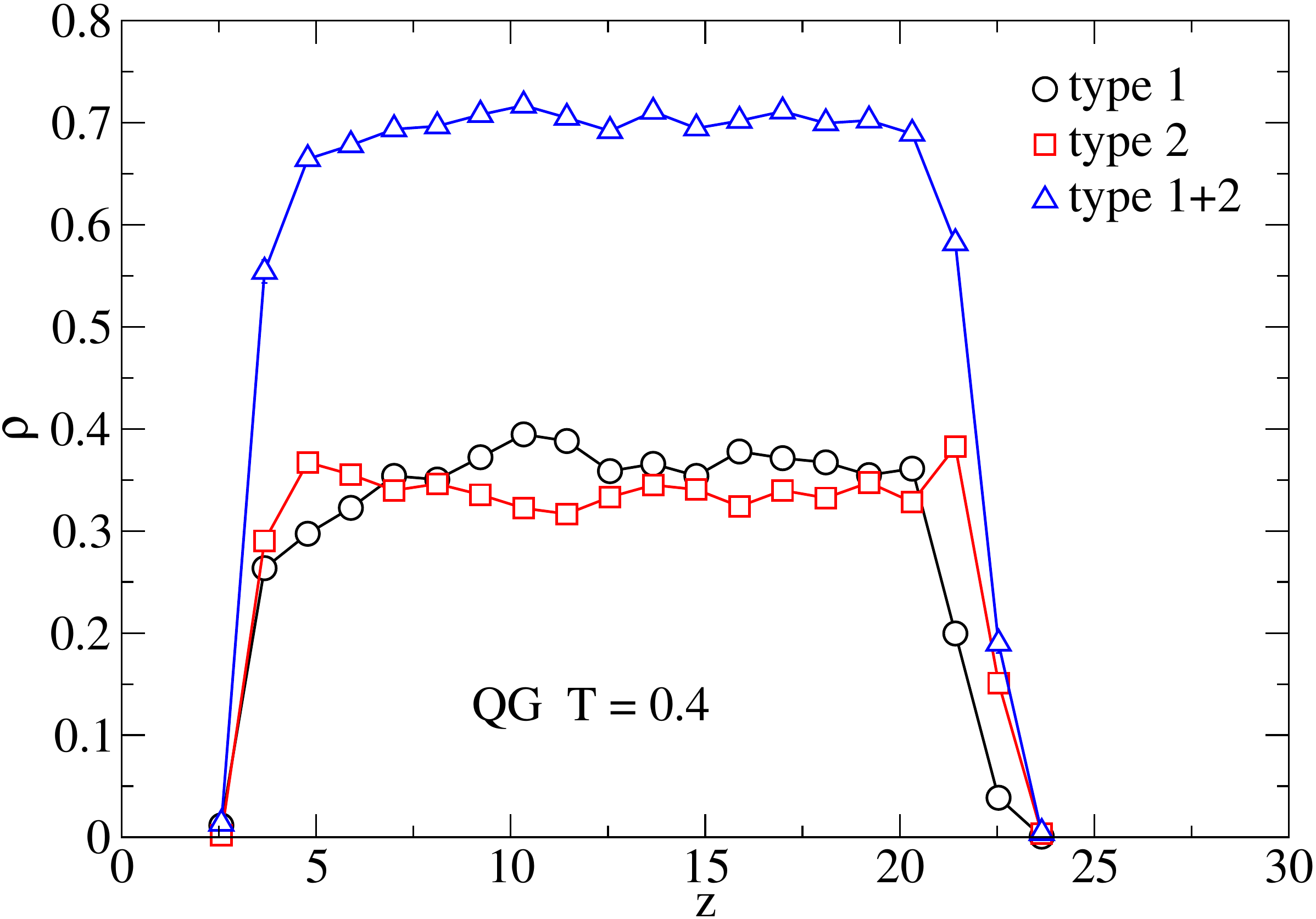}
\includegraphics[width=8cm]{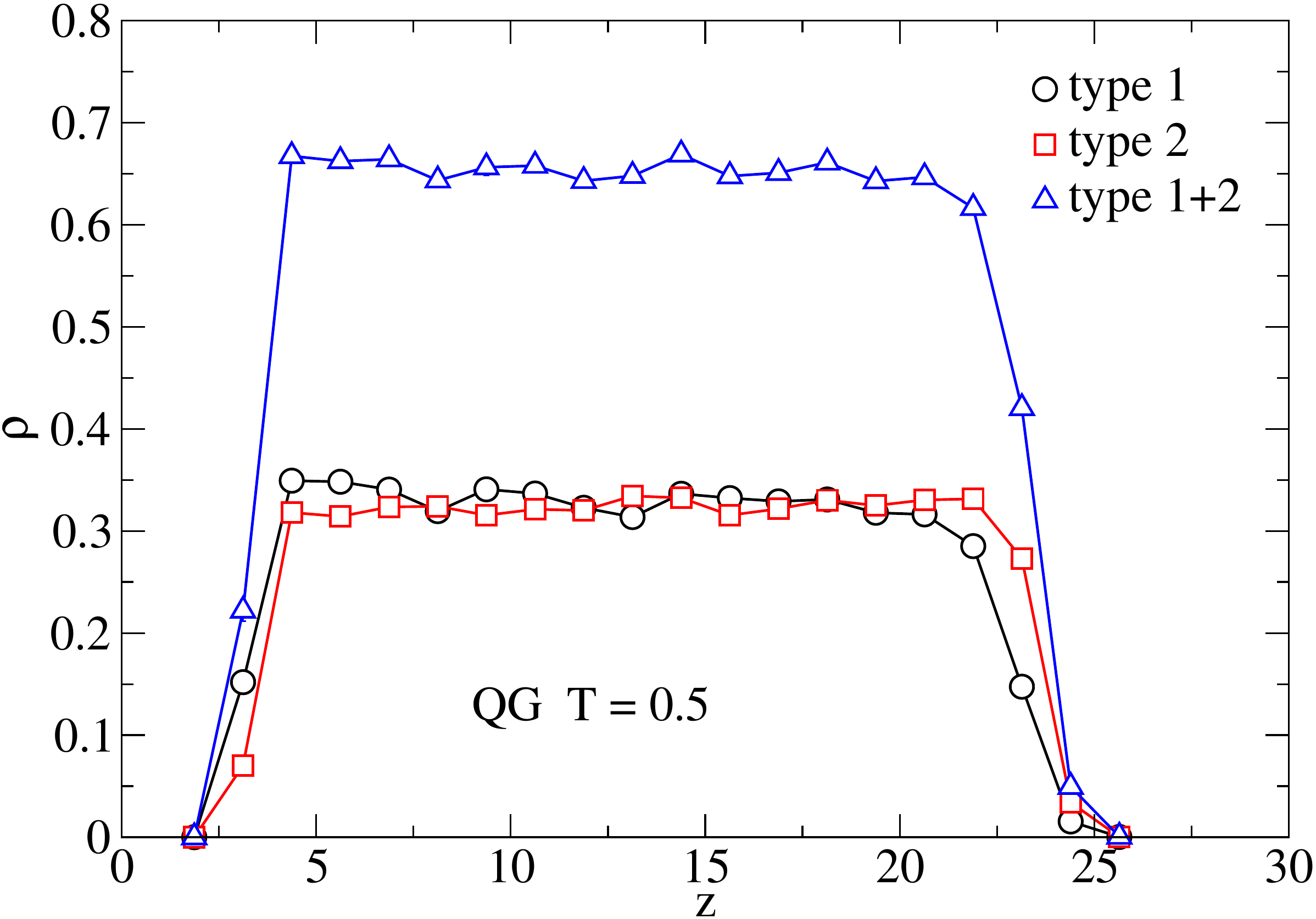}
\caption{\label{fig:rho} Density profile of the deposited layer and annealed layer at temperatures $T=0.4, 0.5$. Density and lengths are in reduced units.}
\end{center}
\end{figure*}

The density profile of the DG and QG glasses as a function of the distance from the bottom of the simulation box $z$ shows a roughly flat distribution of particles (flatter for higher temperatures) in the middle of the layer (we call core in the main text). The density profile sharply goes to zero near the edges (within $\sim 2\sigma$), as shown in Fig.~\ref{fig:rho}.  
The core, in which quantities shown in the main text are computed, is defined as the part of the layer where particles have coordinates $z_{bot}+4\sigma<z<z_{top}-4\sigma$, with $z_{bot}$, $z_{top}$ the bottom and top edges of the layer, respectively. In this way we disregard edge effects related to a change in the density profile and composition of the system.

From the same figure we can see that there is an excess of type 2 particles (with larger diameter respect to type 1) near the free surface. This effect has been observed in previous simulations for deposition of Kob-Andersen (KA) mixtures in 2D \cite{reid2016age} and 3D free-standing layers of KA \cite{Shi2011}. In these works this effect has been attributed to the maximization of type 1 -- type 2 interactions in the layer since for KA mixtures (where type 1 and 2 are commonly indicated with B and A particles, respectively) $\epsilon_{12}>\epsilon_{22}>\epsilon_{11}$ and type 2 and 1 particles are $80\%$ and $20\%$ of the total, respectively. 
In the present case, type 2 and 1 particles are present in the same proportion and all the $\epsilon$ are the same, so that the segregation of type 2 particles we observe at the free surface should be linked to the system minimization of the energetic cost of forming the interface through the maximization of the number, rather than the type, of particle interactions coming from optimal particle packing, an effect that could be relevant also for segregation in the KA model.

\section{Frank-Kasper bonds}

\begin{figure}[h] 
\begin{center}
\includegraphics[width=11cm]{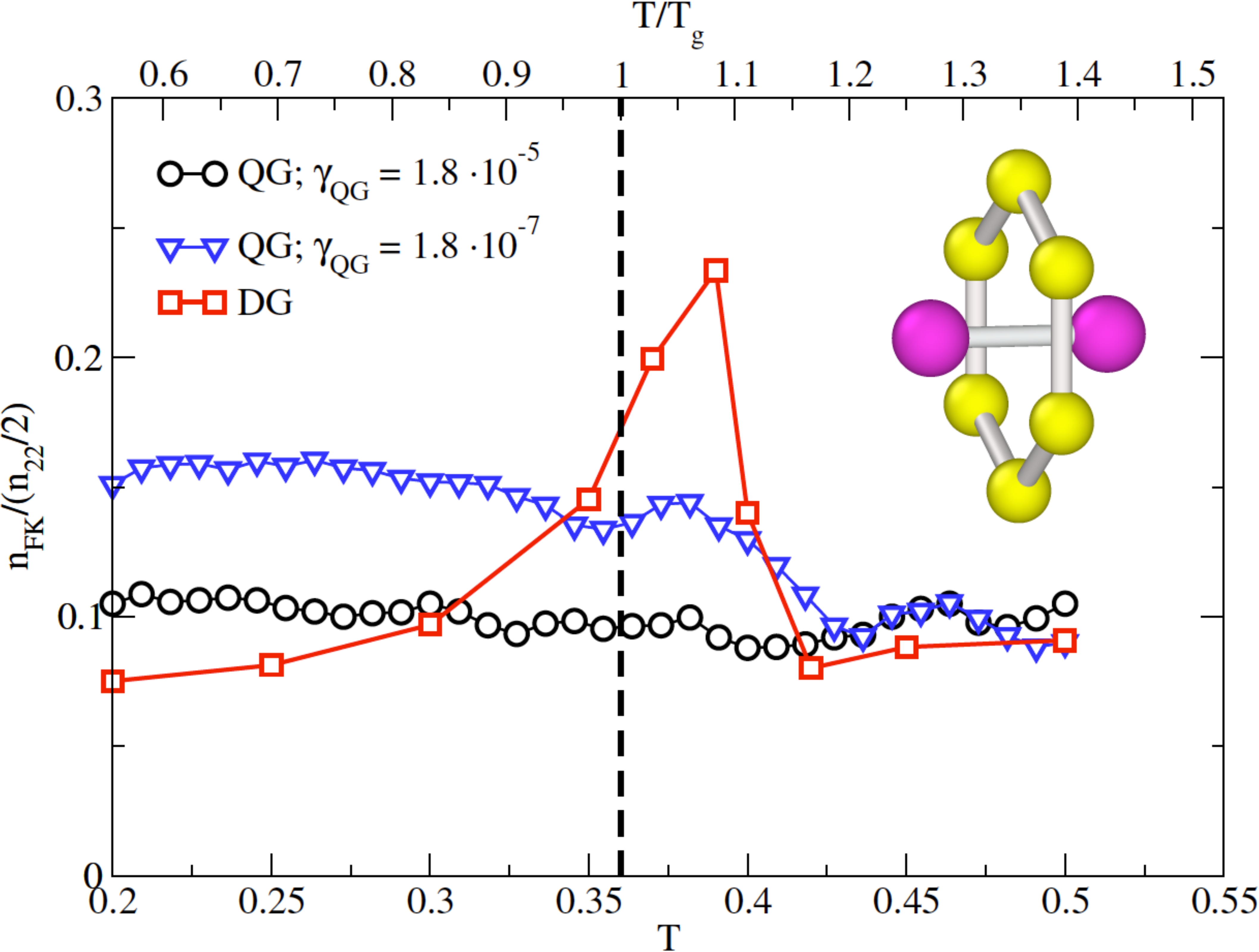}
\caption{\label{fig:FK} Number of Frank-Kasper bonds, $n_{FK}$,  normalized for half the number of particles of type 2, $n_{22}$, versus temperature $T$. USG (red squares) and QG (black circles for $\gamma_{QG}=1.8\cdot 10^{-5}$ and blue triangles for $\gamma_{QG}=1.8\cdot 10^{-7}$).}
\end{center}
\end{figure}

FK bonds connect two neighboring particles of type 2 if they share at least 6 neighbors of type 1 (see snapshot in Fig.~\ref{fig:FK}). We compute neighbors by using the radical Voronoi method suited for additive mixtures.
We repeat for FK bonds versus temperature the same analysis performed with the TCC to find icosahedral order (in Fig.~\ref{fig:TCC2}).
Fig.~\ref{fig:FK} shows that, similarly to icosahedra, FK bonds capture the structural change of USG respect to QG for temperature slightly above $T_g$, as expected considering that FK bonds are involved in the formation of the crystal phase.
%

\section{TCC in Kob-Andersen}

\begin{figure}[h] 
\begin{center}
\includegraphics[width=11cm]{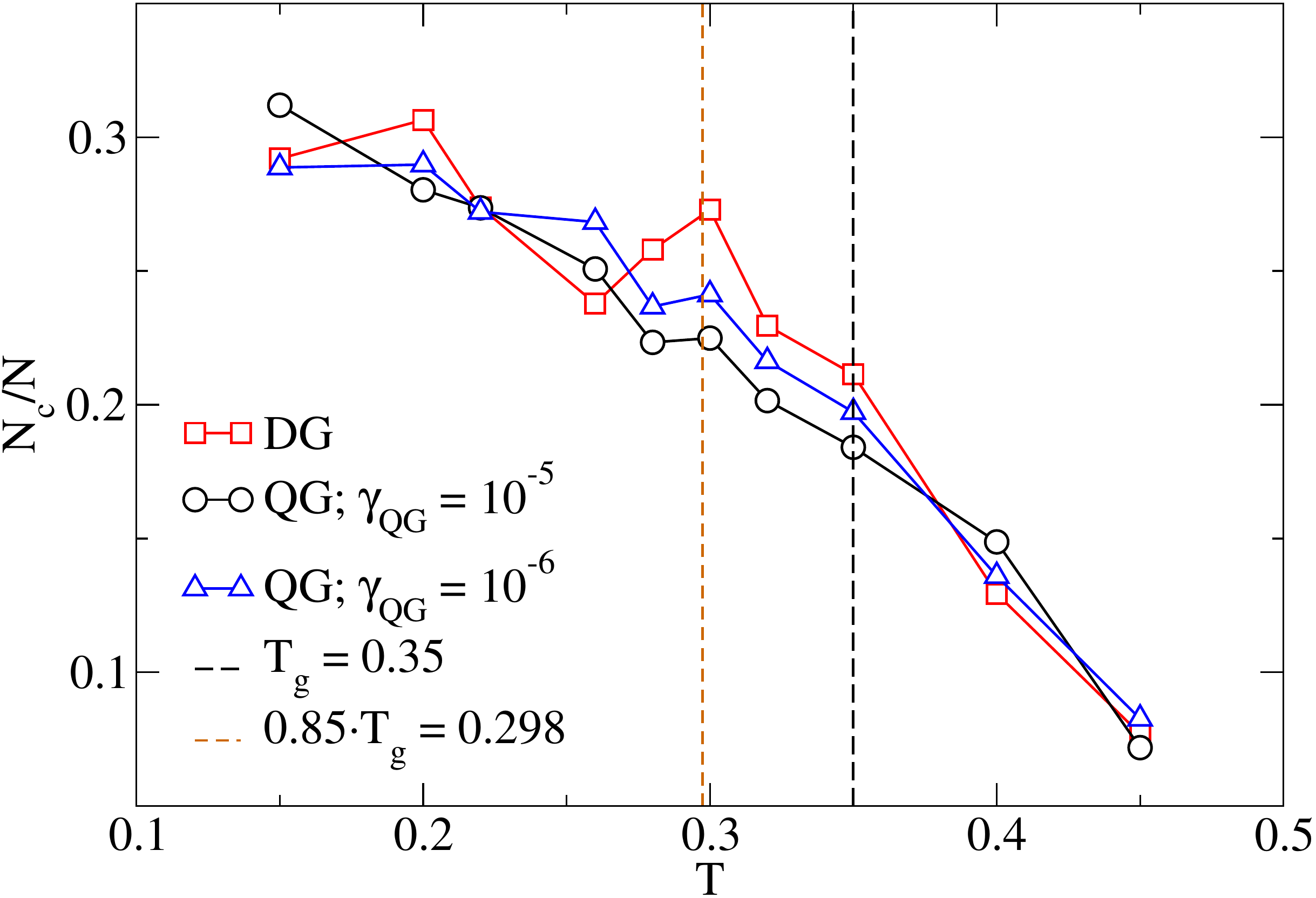}
\caption{\label{fig:KA} Kob-Andersen glassformer. Fraction of particles detected within bicapped square pyramids cluster type (11A motif) with the TCC algorithm. Bond detection method is modified Voronoi with $f_c=1.0$ (see Ref.~\cite{Malins2013}). USG (red squares) and QG (black circles for $\gamma_{QG}=10^{-5}$ and blue triangles for $\gamma_{QG}=10^{-6}$).} 
\end{center}
\end{figure}

In Fig.~\ref{fig:KA} we show the fraction of particles detected within bicapped square pyramids cluster type (11A motif) with the TCC algorithm for the deposited glass and quenched glass for the Kob-Andersen glassformer \cite{Kob1994}.
The substrate is composed of 1000 particles of type A and B with proportion of 8:2, and the deposited layer is composed of 8000 particles with the same proportion.
In reduced units the glass transition temperature is $Tg=0.35$. 
In this case the difference in LFS analysed with the TCC algorithm between the DG and QG is not as striking as for the WAHN glassformer

\newpage

\section{2-point correlation function}

\begin{figure}[h] 
\begin{center}
\includegraphics[width=11cm]{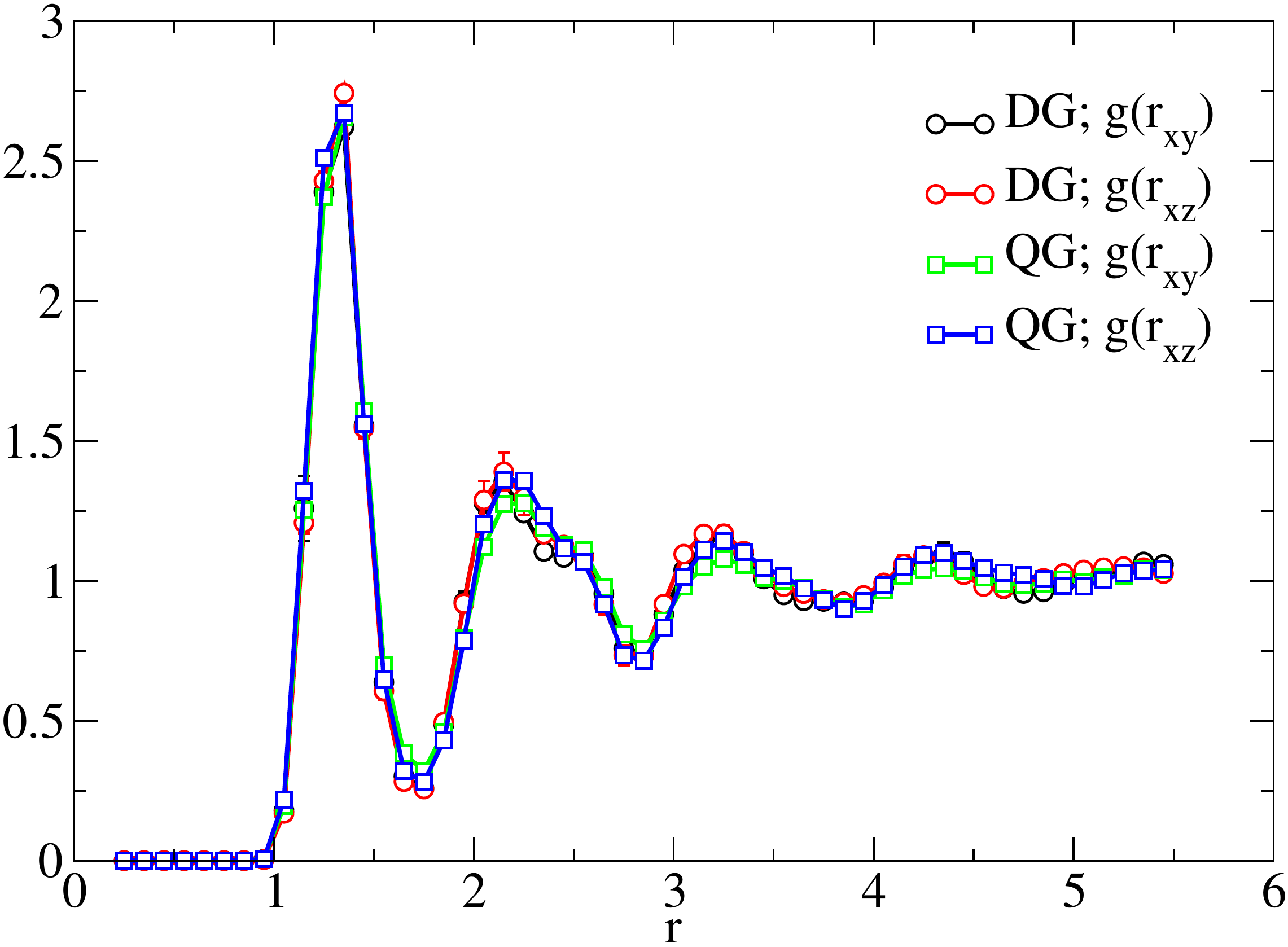}
\caption{\label{fig:gr} 2-point correlation function along the $xy$ and $xz$ plane for the DG and QG with $\gamma_{QG}=10^{-5}$ at $T=0.35$.} 
\end{center}
\end{figure}

\newpage

\section{TCC by layer}

\begin{figure}[h] 
\begin{center}
\includegraphics[width=8cm]{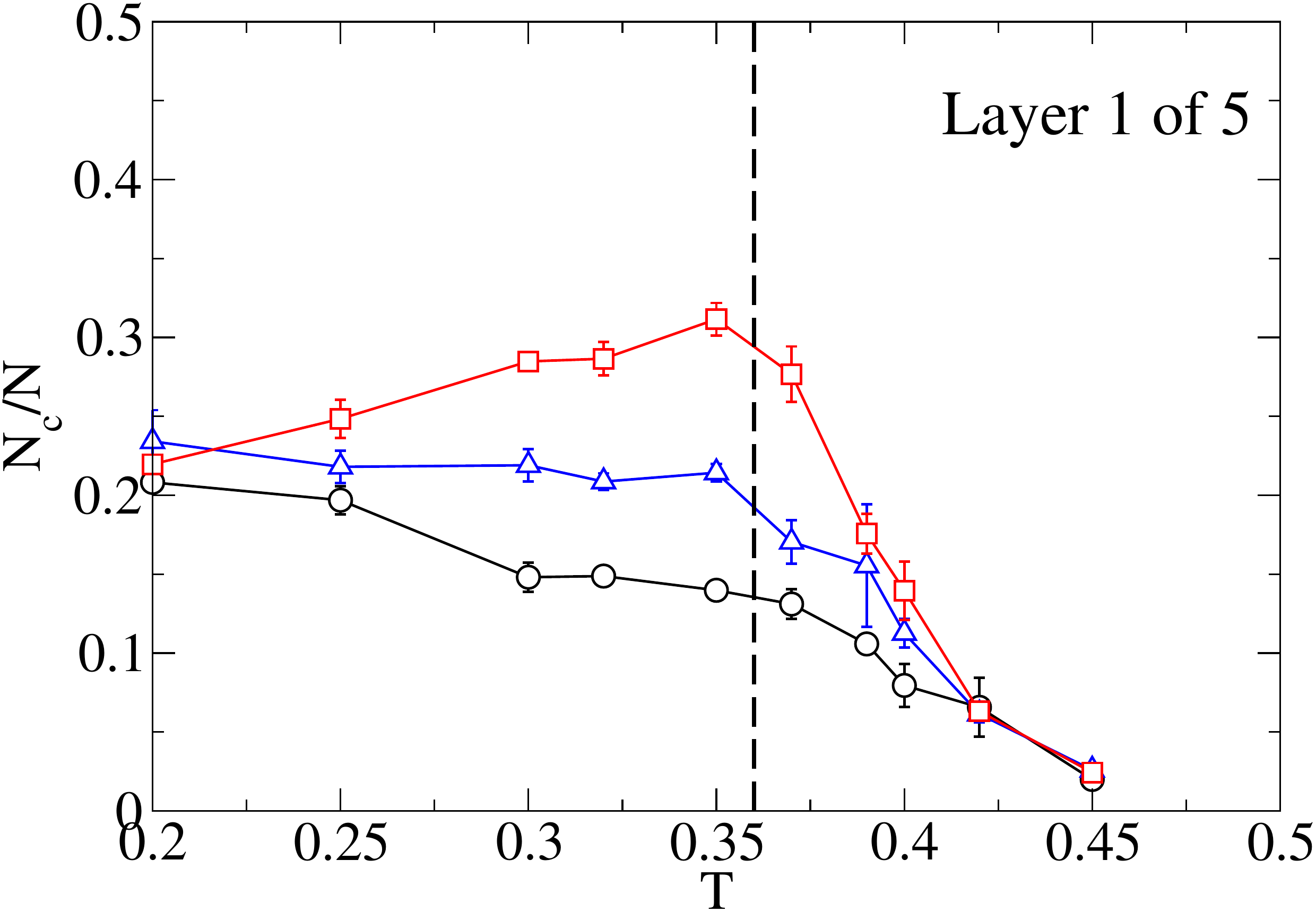}
\includegraphics[width=8cm]{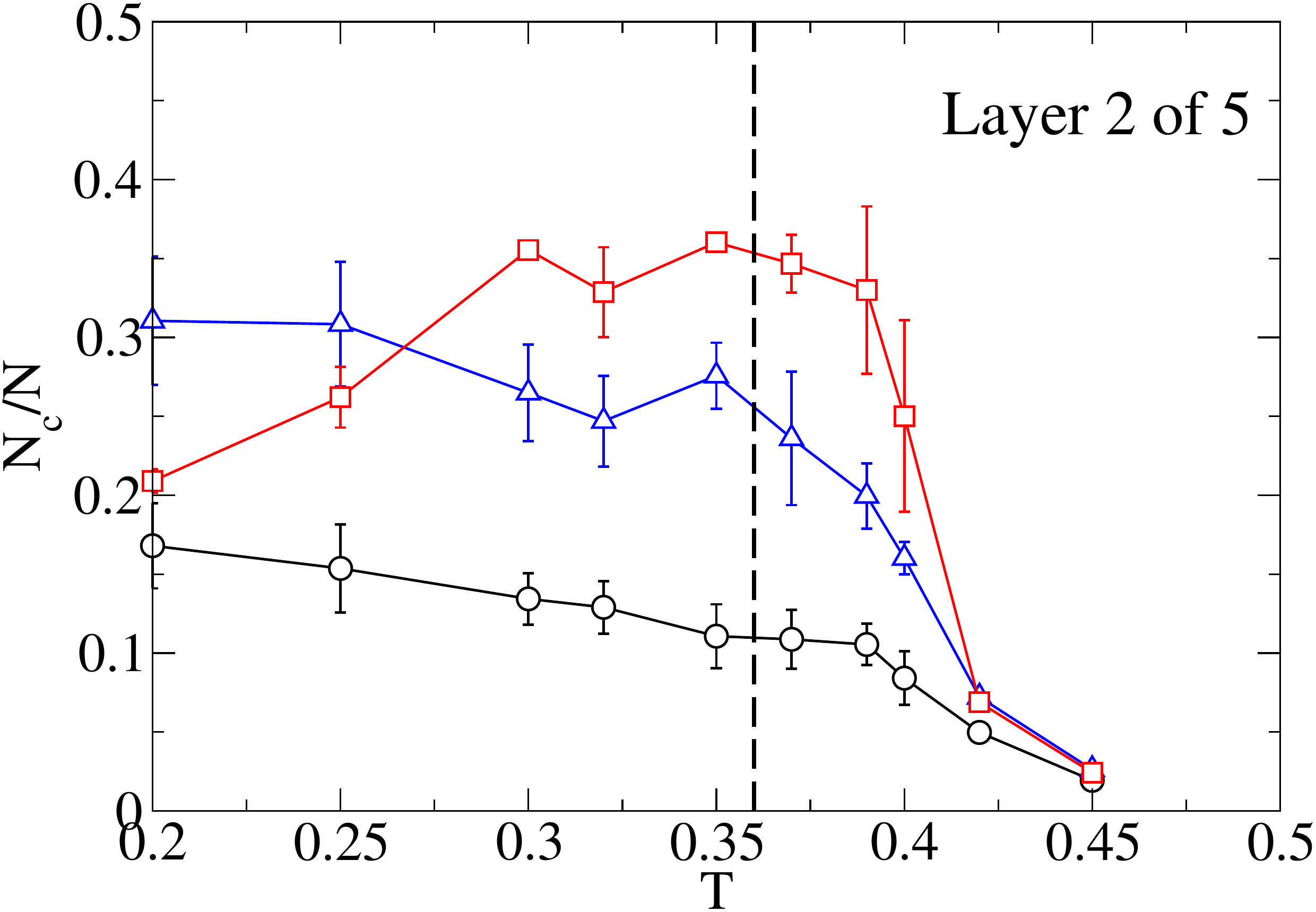}

\includegraphics[width=8cm]{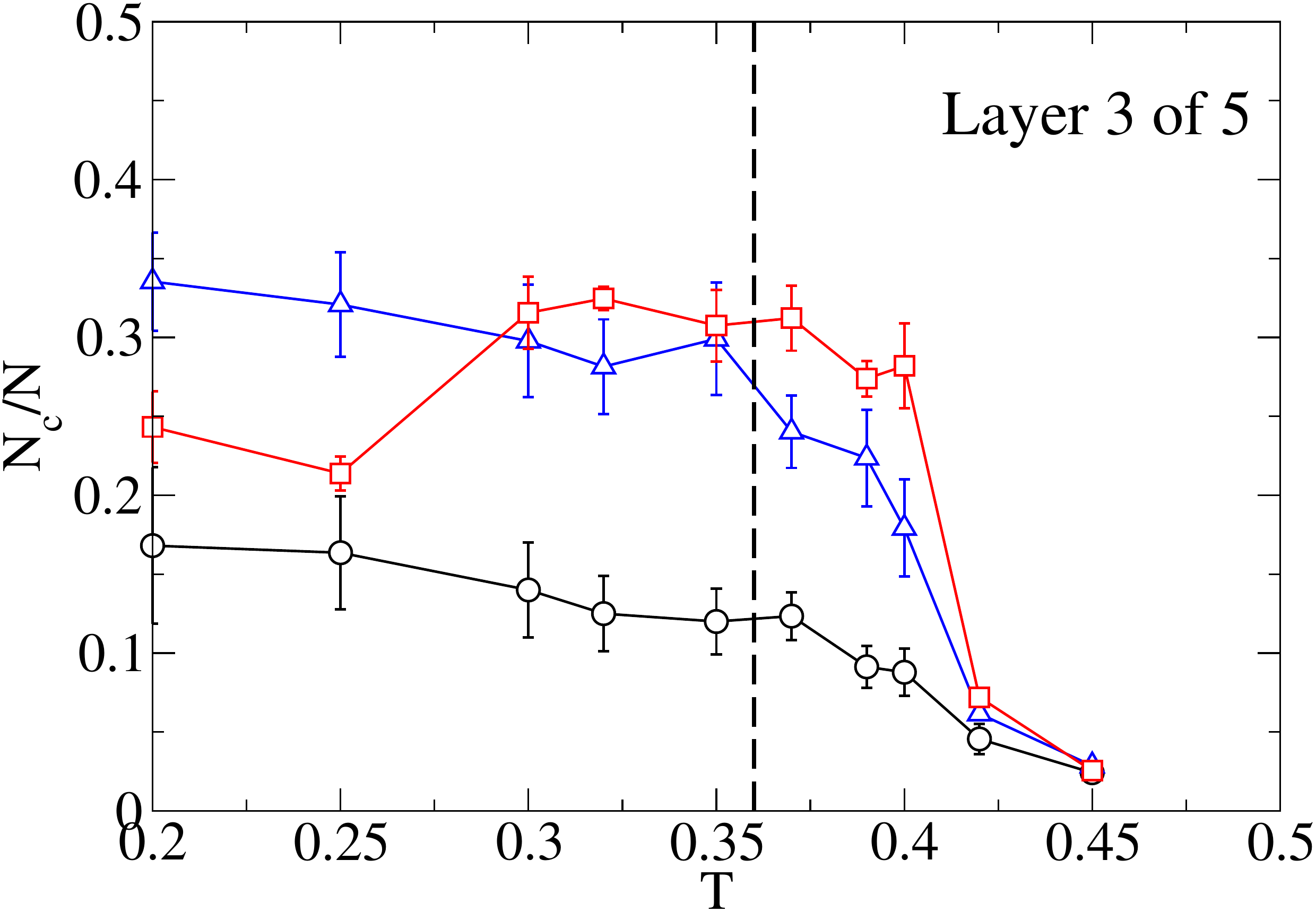}
\includegraphics[width=8cm]{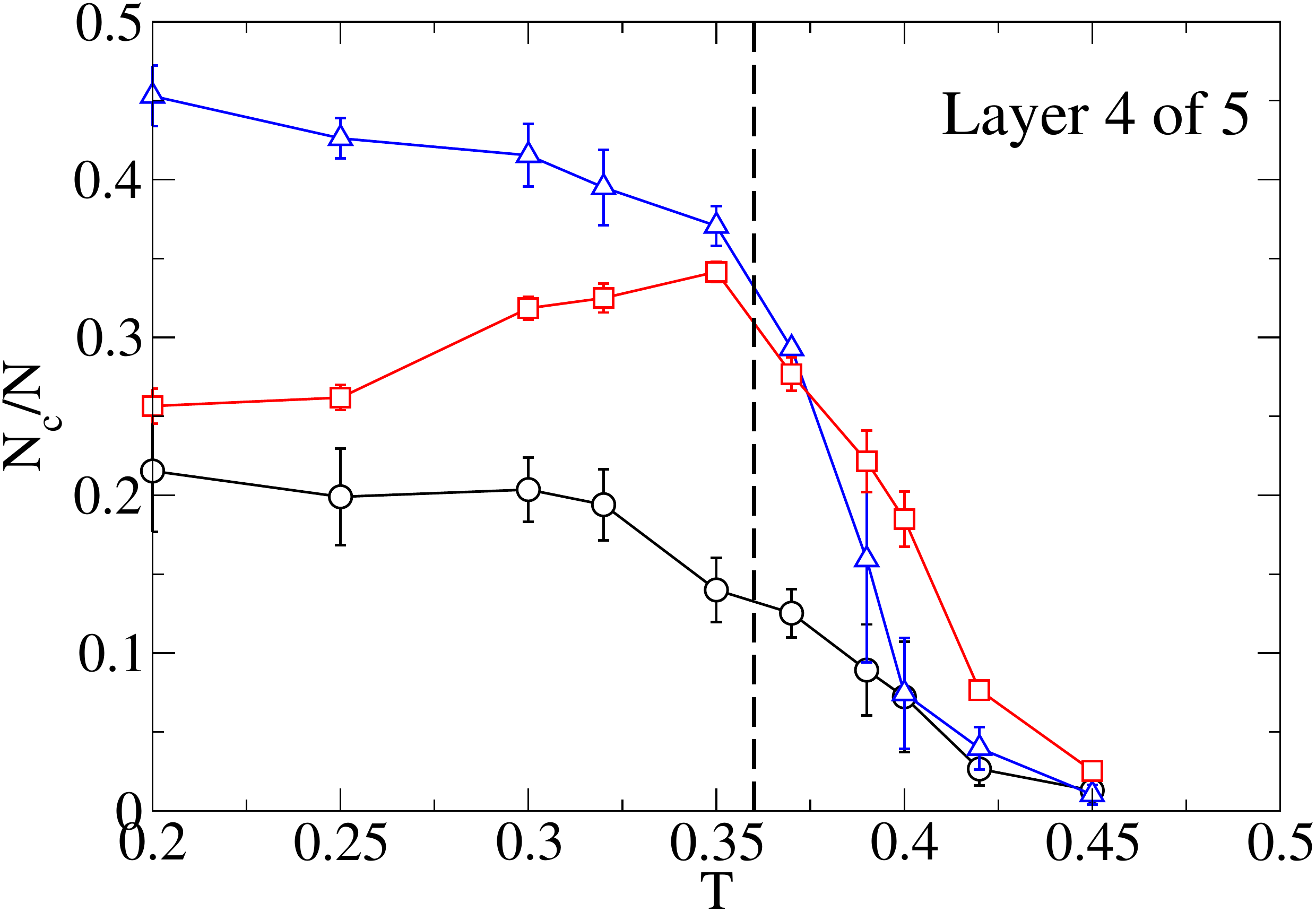}

\includegraphics[width=8cm]{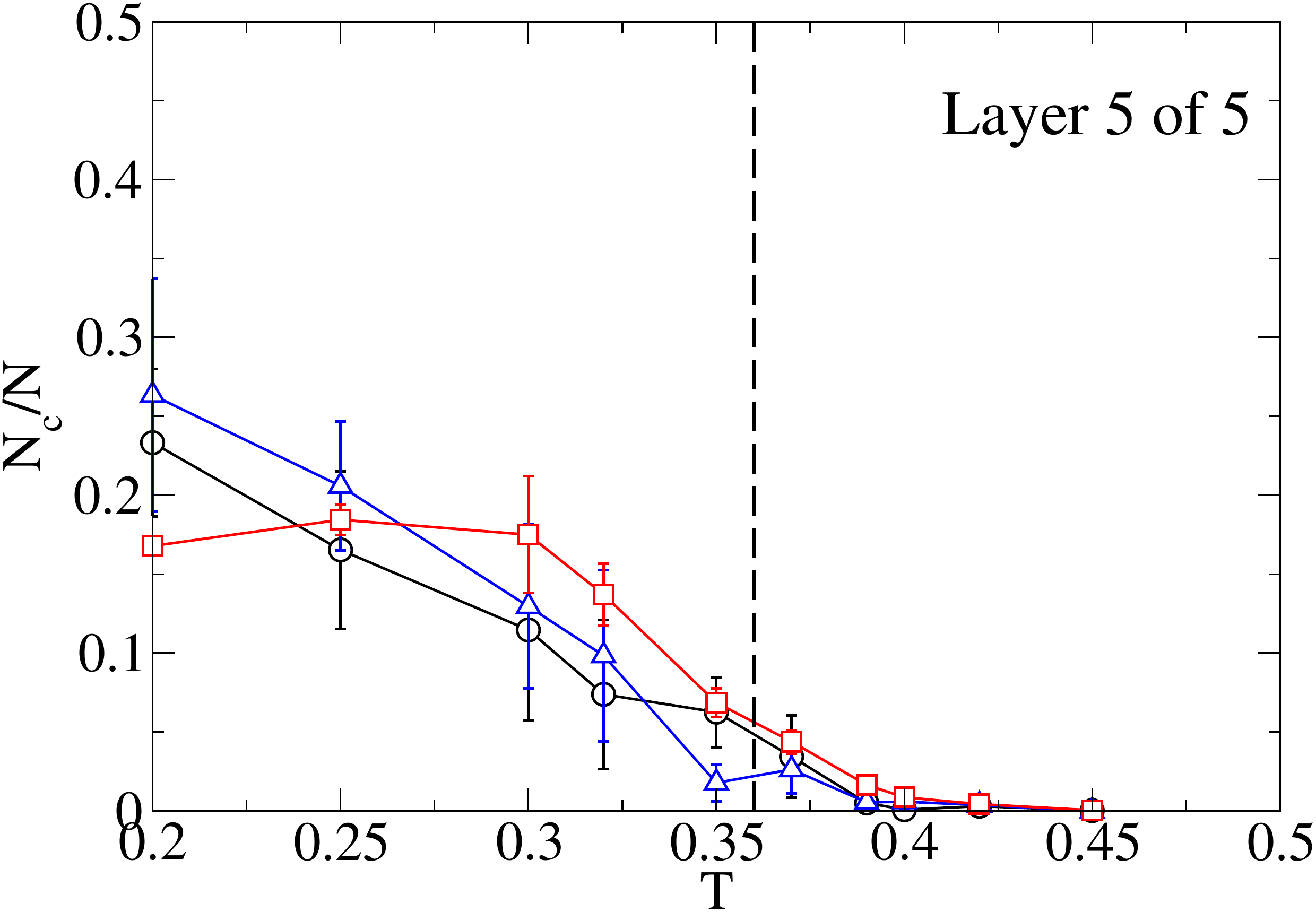}
\includegraphics[width=8cm]{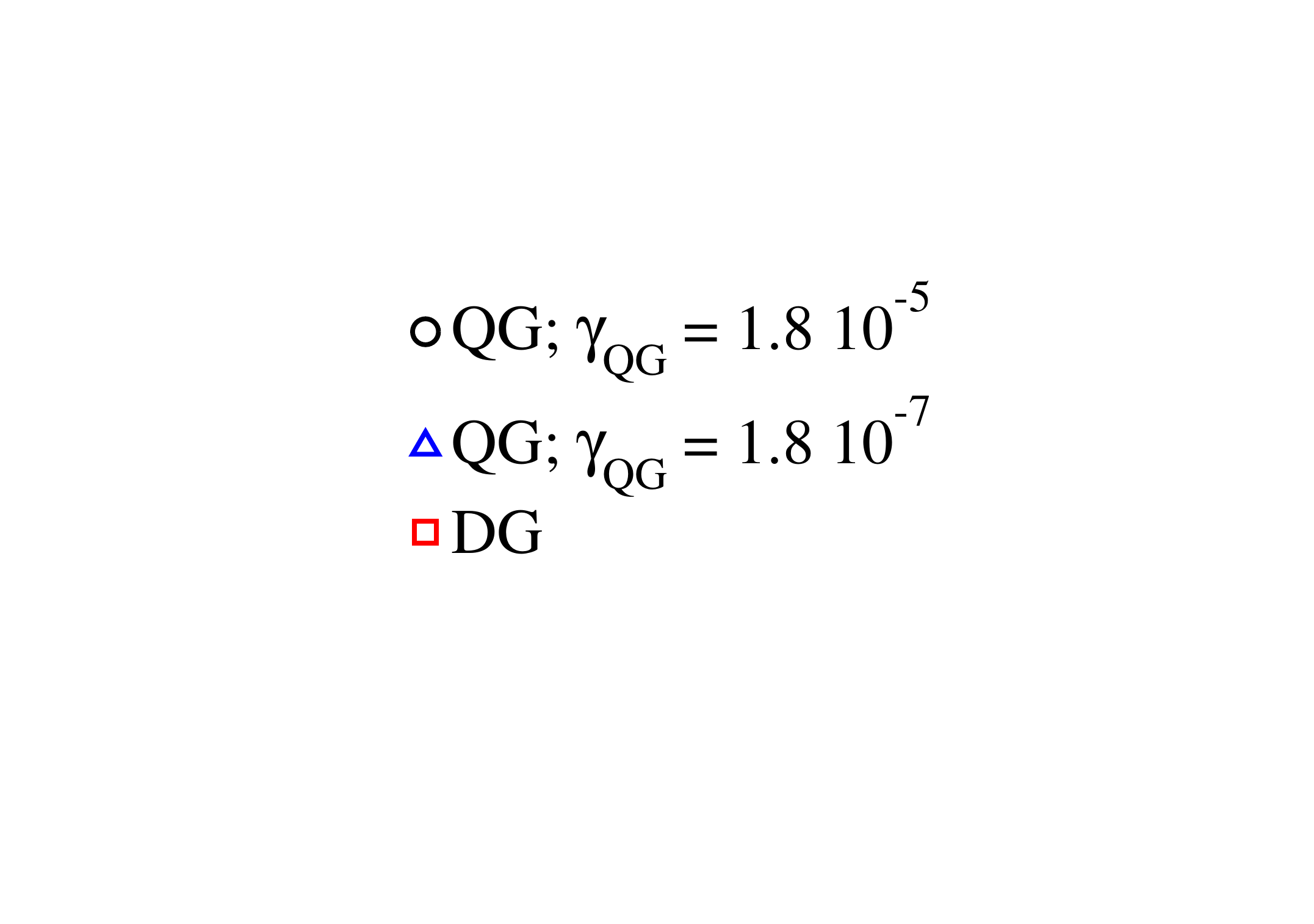}
\caption{\label{fig:tcc_layer} Fraction of particles detected within icosahedra with the TCC algorithm in each one of the 5 regions in which the layer is divided for the DG (red squares), and the QG at two cooling rates (black circles and blue up-triangles).} 
\end{center}
\end{figure}

The TCC is applied to the Wahnstr\"om mixture employing as bond detection method the modified Voronoi with $f_c=0.82$ (see Ref.~\cite{Malins2013}).
From Fig.~\ref{fig:n_c} it is clear that different regions of the layer contribute with a different weight to the LFS content in the core shown in Fig.~\ref{fig:TCC2}. In Fig.~\ref{fig:tcc_layer} we show the fraction of particles detected within icosahedra, $N_c/N$, for 5 different regions into which we divided the layer. From it we can see that $\langle N_c/N\rangle$ for the QG at fast annealing is always lower, for any $T$, with respect to the slow annealing version. From the other hand, at high $T$ the DG has a $\langle N_c/N\rangle$ larger than both QG systems and develops a peak at a temperature $T<T_g$. Further decreasing $T$ the content of $\langle N_c/N\rangle$ in the DG becomes always lower with respect to the QG at slow annealing and in some case also with respect to the QG at fast annealing.

\newpage

\section{2D MSD}

\begin{figure}[h] 
\begin{center}
\includegraphics[width=11cm]{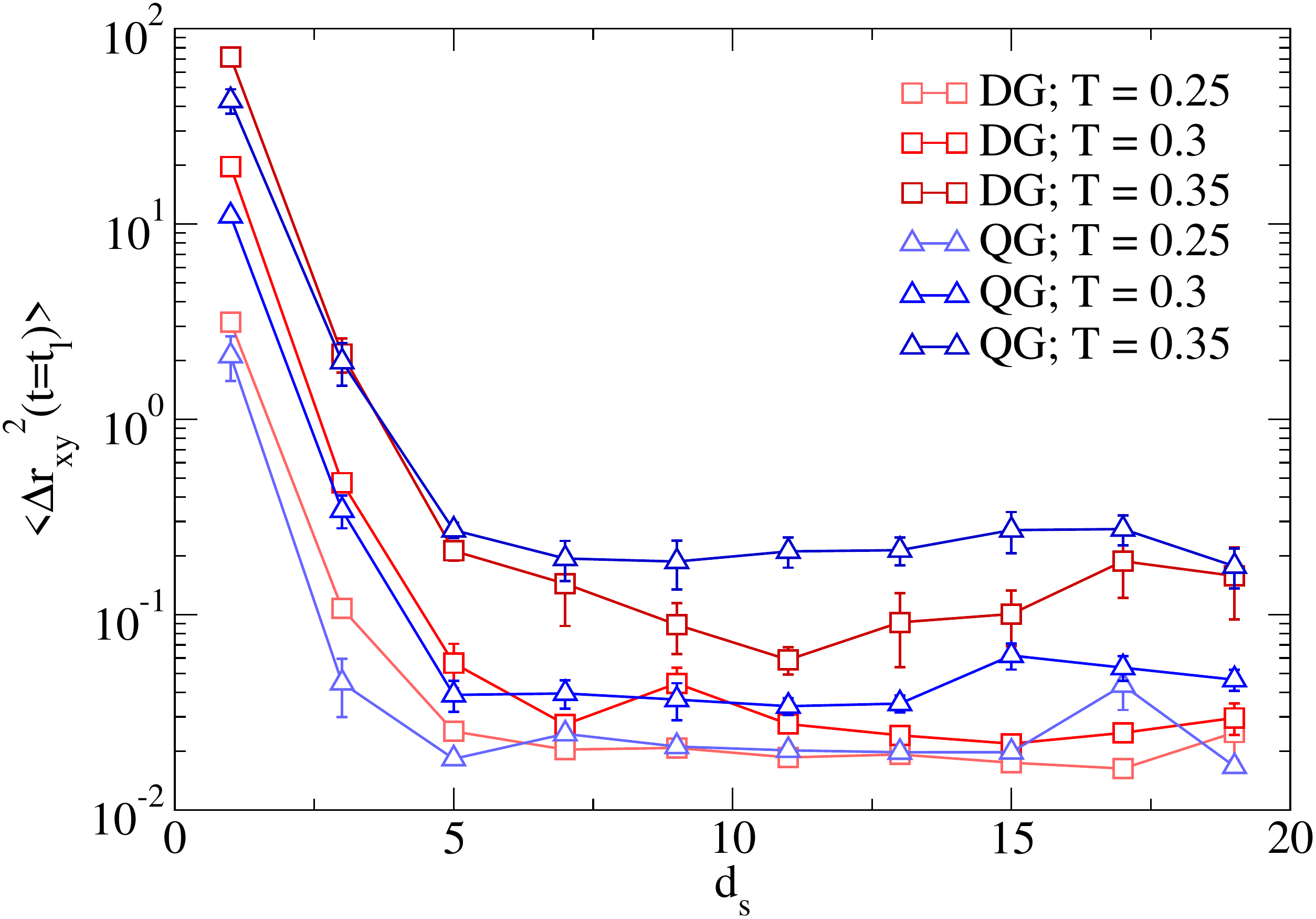}
\caption{\label{fig:msd} MSD in the $xy$ plane at the lag time $t_l=2\cdot 10^3$ as a function of the distance from the free surface $d_s$ for the DG and QG with $\gamma_{QG}=1.8\cdot 10^{-7}$ at different temperatures.} 
\end{center}
\end{figure}

\begin{figure}[h] 
\begin{center}
\includegraphics[width=11cm]{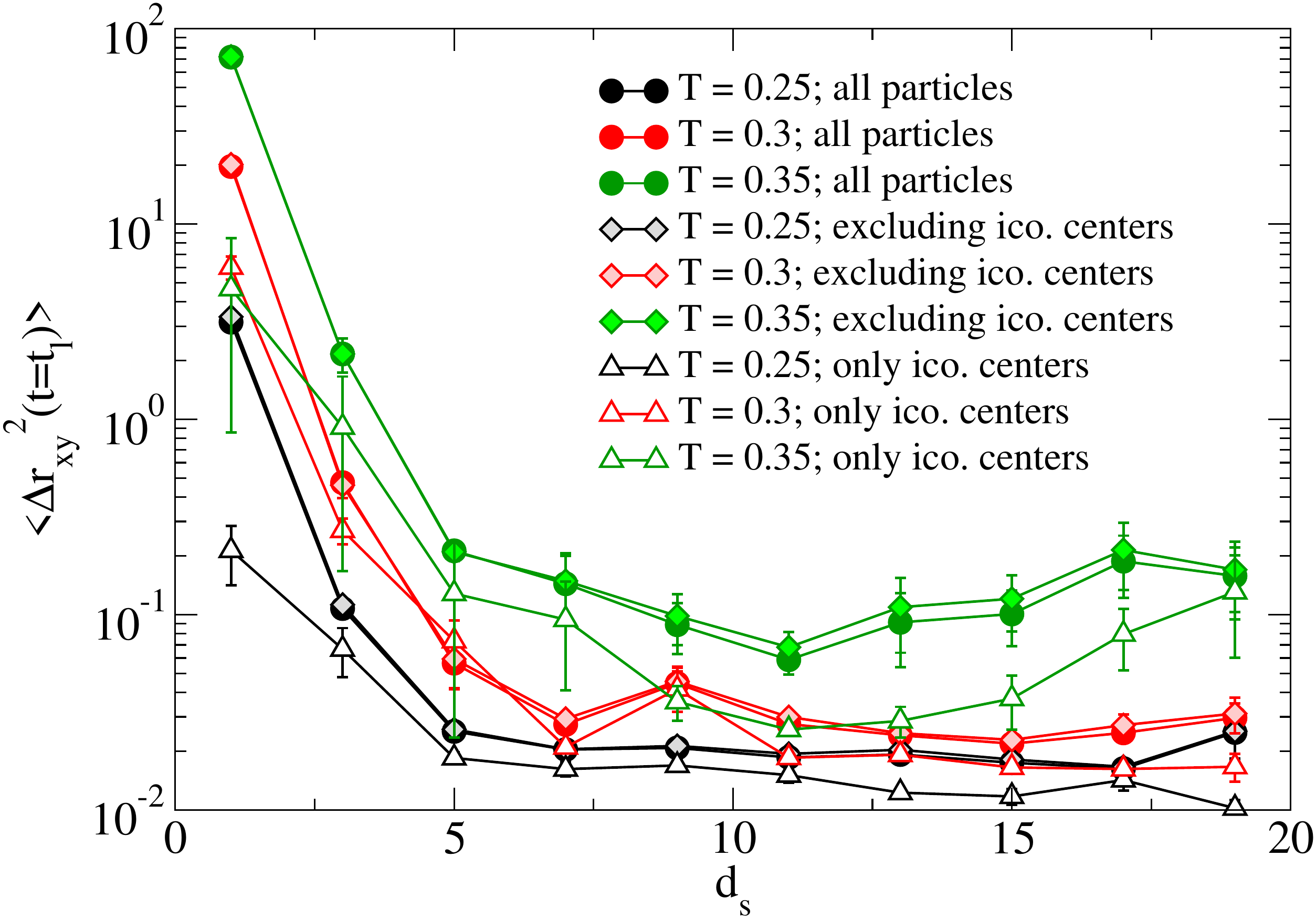}
\caption{\label{fig:msd2} MSD in the $xy$ plane at the lag time $t_l=2\cdot 10^3$ as a function of the distance from the free surface $d_s$ for the DG at different temperatures and different computational methods, as specified in the main text and in the legend where ico. stands for icosahedra.} 
\end{center}
\end{figure}

From Fig.~\ref{fig:msd}, the MSD in the core of the layer is lower for the DG with respect to the QG at temperatures corresponding to its higher stability (and higher content in LFS, see Fig.~\ref{fig:n_c}), in agreement with the results found in Ref.~\cite{zhang2022}. By computing the 2D MSD for particles that never become icosahedra centers and only for these last one, we see (Fig.~\ref{fig:msd2}) a correlation between dynamics and structural classification, i.e., icosahedra are slower.

\newpage

\section{\mbox{\bf{$\Delta u_{core}$}}}

\begin{figure}[h] 
\begin{center}
\includegraphics[width=12cm]{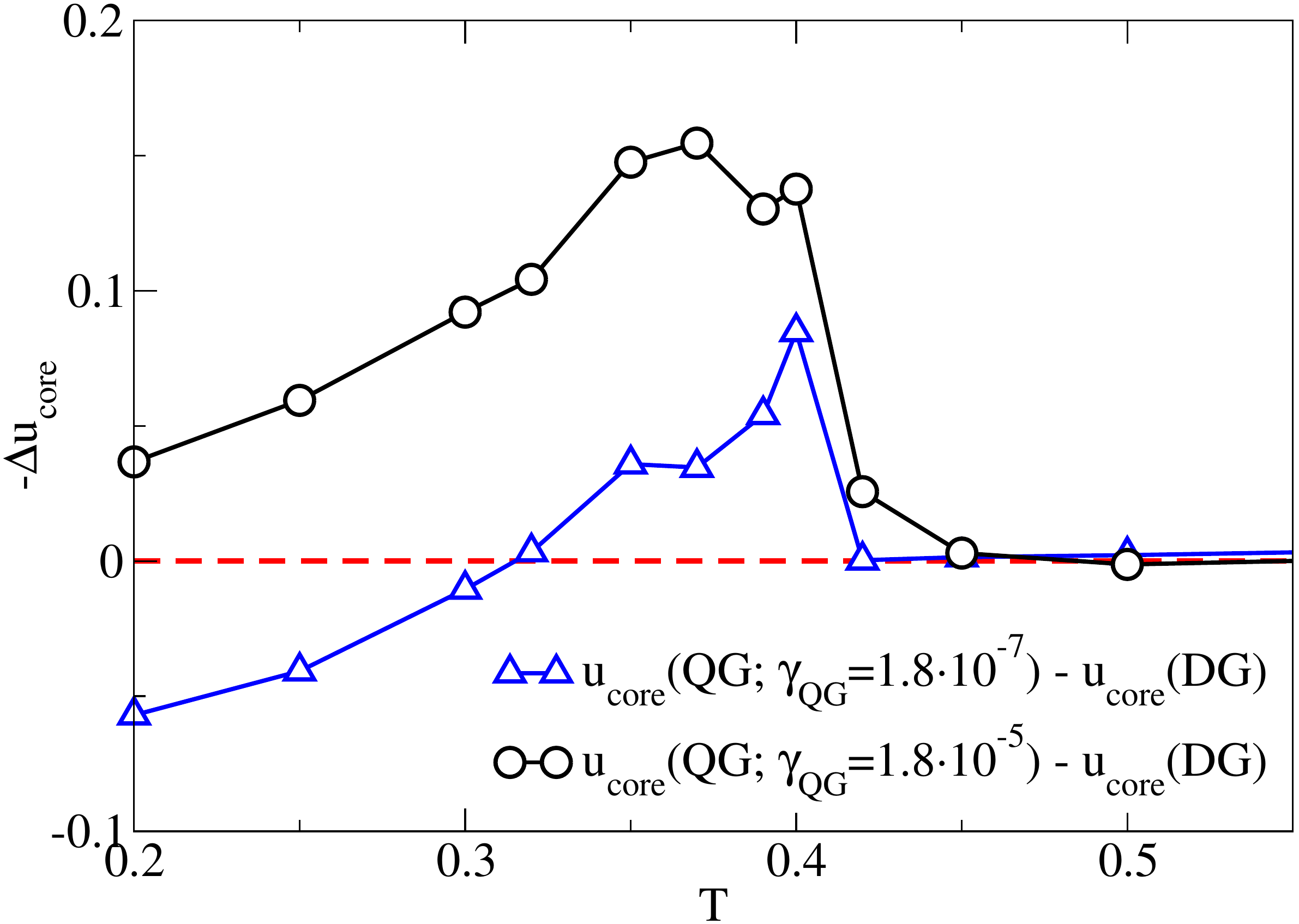}
\caption{\label{fig:Delta_u_SM} $u_{core}$ difference between QG at $\gamma_{QG}=1.8\cdot 10^{-7}$ and DG.} 
\end{center}
\end{figure}

\newpage

\section{Density profile of the fraction of particles detected within icosahedra}

\begin{figure}[h] 
\begin{center}
\includegraphics[width=12cm]{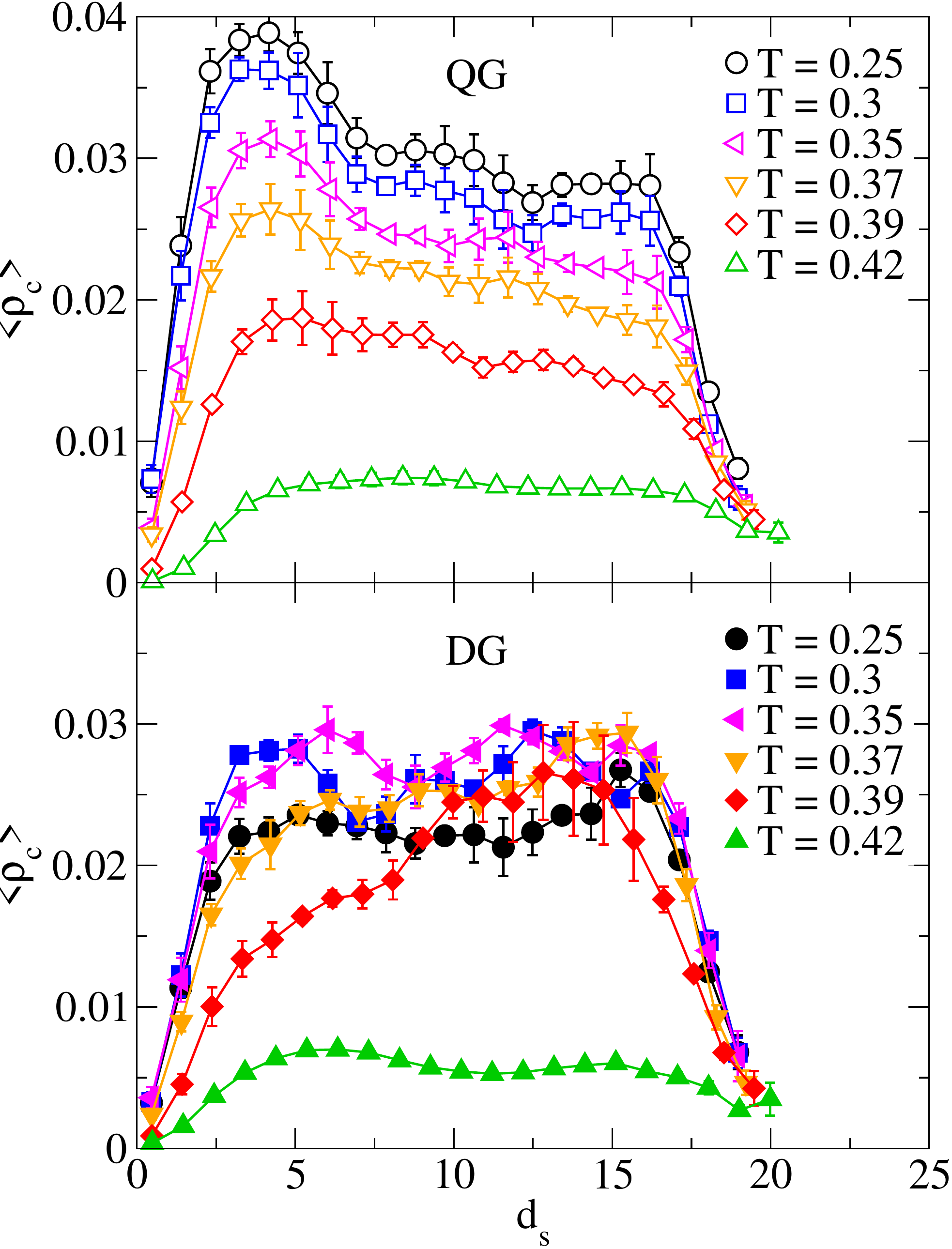}
\caption{\label{fig:Fig4_SM} Average density profile of the
fraction of particles detected within icosahedra, $\langle\delta N_c/N\rangle$, as a function
of the distance from the free surface $d_s$ for the QG (upper panel) and DG (lower panel) at different $T$.} 
\end{center}
\end{figure}

\end{document}